\documentclass[aapm,graphicx]{revtex4-1}
\pdfoutput=1
\draft
\setcitestyle{super}
\usepackage[sort&compress]{natbib}
\usepackage{amsfonts}
\usepackage{amssymb}
\usepackage{amsmath,epsfig}
\usepackage{subfigure}
\usepackage{fmtcount}
\usepackage{array}
\usepackage{rotating}
\usepackage{fmtcount}
\usepackage{mathbbol}
\usepackage{dsfont}
\usepackage{placeins}
\usepackage{booktabs,xcolor,siunitx}
\usepackage{slashbox}
\usepackage{url}
\usepackage[caption=false]{subfig}
\usepackage{textcomp}

\usepackage{multirow}

\usepackage{threeparttable}
\usepackage{nomencl}
\usepackage{soul}
\usepackage{color}
\usepackage{algorithm}
\usepackage{algorithmic}
\usepackage{subfigure}
\usepackage{setspace}
\usepackage{algorithm}
\usepackage{algorithmic}

\makenomenclature

\usepackage{makeidx}
\makeindex 

\begin{document}
\def\vector{\underline}
\def\matrix{}
\renewcommand{\vec}[1]{\mathbf{#1}}
\newcommand{\mat}[1]{\mathbf{#1}}

\title{Reference-Based MRI}

\author{Lior Weizman\footnote{Corresponding author, e-mail:weizmanl@tx.technion.ac.il, phone:+972-4-8291724}}

\address{Department of Electrical Engineering, Technion - Israel Institue of Technology, Israel}
\author{Yonina C. Eldar}

\address{Department of Electrical Engineering, Technion - Israel Institue of Technology, Israel}

\author{Dafna Ben Bashat}

\address{Functional Brain Center, Tel Aviv Sourasky Medical Center, Sackler Faculty of Medicine and Sagol School of Neuroscience,  Tel Aviv University, Israel}

\date{\today}

\begin{abstract}
{\bf Purpose:} In many clinical MRI scenarios, existing imaging information can be used to significantly shorten acquisition time or to improve Signal to Noise Ratio (SNR).  In this paper the authors present a framework for fast MRI by exploiting a reference image (FASTMER).

{\bf Methods:} The proposed approach utilizes the possible similarity of the reference image that exists in many clinical MRI imaging scenarios. Such scenarios include similarity between adjacent slices in high resolution MRI, similarity between various contrasts in the same scan and similarity between different scans of the same patient. The authors take into account that the reference image may exhibit low similarity with the acquired image and develop an iterative weighted approach for reconstruction, which tunes the weights according to the degree of similarity.

{\bf Results:} Experimental results demonstrate the performance of the method in three different clinical MRI scenarios: SNR improvement in high resolution brain MRI, exploiting similarity between T2-weighted and fluid-attenuated inversion recovery (FLAIR) for fast FLAIR scanning and utilizing similarity between baseline and follow-up scans for fast follow-up. Results outperform reconstruction results of existing state-of-the-art methods.

{\bf Conclusions:} The authors present a method for fast MRI by exploiting a reference image. The method is based on an iterative reconstruction approach that supports cases in which similarity to the reference scan is not guaranteed, which enables the applicability of the method to a variety of MRI applications. Thanks to the existence of reference images in various clinical imaging scenarios, the proposed framework can play a major part in improving reconstruction in many MR applications.
\end{abstract}



\maketitle

\newpage

\section{Introduction}
\label{Introduction} 
Magnetic resonance imaging (MRI) data is sampled in the spatial Fourier transform (a.k.a. k-space) of the object under investigation. In many cases, the k-space is sampled below the Nyquist rate due to constraints in the implementation
of the k-space trajectory that control the sampling pattern (e.g., acquisition duration and smoothness of gradients). Mostly, prior assumptions on the nature of the data are taken into account in the reconstruction process, to overcome imaging artifacts due to insufficient sampling. We can roughly divide MRI reconstruction approaches from undersampled k-space into two families: single- and multiple-image based recovery.

The first family of methods exploits prior assumptions on a single MR image, in order to improve its reconstruction from undersampled data. Since the introduction of Compressed Sensing (CS) \cite{candes2006compressive,donoho2006compressed,eldar2012compressed,eldar2015sampling} to the field of MRI \cite{lustig2007sparse}, many MRI reconstruction approaches exploit the fact that MR images are highly compressible, by formalizing the image reconstruction problem as an $\ell_1$ minimization problem. Wavelet transform sparsity has been widely used as a sparsifying transform for brain MRI. Total Variation (TV) has been used for MR images which are sparse in the image domain, such as angio-MRI \cite{lustig2007sparse,ma2008efficient}. Other approaches focus on learning the sparsifying transform or using a dictionary developed exclusively for MRI \cite{ravishankar2011mr,qu2012undersampled,caballero2014dictionary}. 

The second family of techniques exploit similarity to a single reference image or within a series of MRI images. In Table \ref{table1} we present a concise review of prior art in methods that exploit reference images to speed-up acquisition or to improve image reconstruction. Most algorithms were applied to dynamic or Diffusion MRI, where multiple images are acquired at a single imaging session. This allows the exploitation of similarity along the temporal dimension, assuming that only parts of the field-of-view (FOV) change over time. State-of-the-art approaches exploit the temproal similarity in a generalized reconstruction scheme \cite{liang1991generalized,hanson1996fast,hass1999maximum,yun2009high,nguyen2013modified}, in a Bayesian-based approach \cite{haldar2013improved}, in a CS framework \cite{lustig2006kt,lang2008accelerating,gamper2008compressed,jung2009k,chen2011framework}, or by enforcing low-rank reconstruction \cite{chiew2014k,adluru2015mri,otazo2015low}.

Additional works focus on structural MRI. Some use a-priori information to improve reconstruction of single-contrast MRI \cite{wu2011prior,peng2011reference,du2012compressed,lam2011motion}. In multiple-contrast MRI, structural similarity between different contrasts in the same scan is assumed and can be used to enhance reconstruction \cite{bilgic2011multi,qu2013magnetic,qu2014magnetic,huang2014fast}. A few works have been published on exploiting similarity in longitudinal MRI, where images were acquired at different time points \cite{tsao2001unifying,samsonov_ismrm,li2015incorporation}. Finally, the use of reference images to improve reconstruction has also been tested in other imaging modalities, such as Magnetic Resonance Spectroscopy (MRS) \cite{hu1988slim,liang1991generalized}, Positron Emission Tomography (PET) \cite{gindi1993bayesian,ouyang1994incorporation,somayajula2011pet} and X-Ray Computed Tomography (X-Ray CT) \cite{chen2008prior}.

Taking a deeper look into the rightmost column of Table \ref{table1}, we observe that multiple-image based reconstruction is application specific; since similarity between multiple images takes on different forms, a separate reconstruction approach was developed for each MRI application, exploiting its specific nature. No general sampling and reconstruction scheme which fits a variety of multiple-image MRI applications has been developed so far. Moreover, most of the methods rely on the assumption that there is substantial similarity between the images in the series (e.g. dynamic imaging), in the image or in some transform domain. Assuming similarity of intensity between the reference image and the current scan may not always be valid (e.g. when a different imaging contrast is used as a reference or due to misalignment between images) and therefore may lead to undesired reconstruction results.

Recently, we introduced an iterative approach for sampling and reconstruction exploiting similarity across multiple MRIs of the same patient \cite{weizman2014application}. We considered the acquisition of a follow-up MRI, given the baseline scan of the same patient. We took into account that baseline and follow-up images may not exhibit similarity and developed an iterative weighted  mechanism that adjusts the reconstruction parameters and the sampling locations during real-time scanning. While the idea of optimizing the sampling locations has also been proposed by others \cite{ravishankar2011adaptive}, it suffers from practical difficulties due to the necessity to solve many computationally heavy $\ell_1$ minimization problems during the acquisition process.

In a related work \cite{weizman2014exploiting}, we have shown that enforcing similarity between adjacent, low SNR, thin MRI slices, can lead to significant improvement in SNR that obviates the need for multiple excitations for high SNR. Our approach is based on modifying the RF pulse signal for different weighting of the thin slices in the acquired data, which might be difficult in practical implementation in existing MRI hardware. 

In a recent conference paper we presented the initial concept of applying a reference-based approach for multiple MRI applications \cite{weizman2015fast}. The conference paper includes an earlier version of the model described in this paper, with preliminary results obtained via retrospective 2-dimensional random sampling.



\begin{center}
\begin{table*}
\scriptsize
   \caption{Summary of related methods that exploit a reference image for fast MRI. Our proposed method is the last in the table. Abbreviations used in the table are: DCE: dynamic contrast enhanced, CS: compressed sensing, DWI: diffusion weighted imaging, TV: total variation, fMRI: functional MRI}
    \begin{tabular}{|>{\raggedright\arraybackslash}p{3cm} | >{\raggedright\arraybackslash}p{10cm} | >{\raggedright\arraybackslash}p{4cm} | }
     \hline
     \textbf{Author} & \textbf{Description} & \textbf{Imaging Application Tested}\\ \hline
     Liang et al. \cite{liang1994efficient} & Exploiting temporal similarity in dynamic MRI using generalized scheme imaging & Dynamic MRI (dynamic T1-weighted and diffusion MRI) \\ \hline
     Hanson et al. \cite{hanson1996fast} & Exploiting two high resolution reference images to improve dynamic imaging in a generalized scheme & Dynamic MRI (DCE MRI) \\ \hline
     Hess et al. \cite{hass1999maximum} & Exploiting reference image for generation of basis functions, used to improve dynamic MRI & Dynamic MRI (MR angiography) \\ \hline
     Tsao et al. \cite{tsao2001unifying} & Incorporating reference image and prior on changed regions for improved reconstruction & Longitudinal MRI \\ \hline
     Tsao et al. \cite{tsao2003k} & Exploiting spatiotemporal correlations for dynamic MRI (training-based approach) & Dynamic MRI (cardiac imaging)\\ \hline
     Lustig et al. \cite{lustig2006kt} & Random sampling in k-t space, reconstruction based on wavelet-Fourier sparsity & Dynamic MRI (cardiac imaging) \\ \hline
     Haldar et al. \cite{haldar2008anatomically} & Using anatomical priors to improve SNR via penalized ML   & Single-contrast MRI \\ \hline
     Lang et al. \cite{lang2008accelerating} & Exploiting similarity to a reference image in a CS framework & Dynamic MRI (brain DCE) \\ \hline
     Gamper et al. \cite{gamper2008compressed} & Exploiting sparsity in the x-f space for dynamic MRI & Dynamic MRI (cardiac imaging) \\ \hline
     Jung et al. \cite{jung2009k} & Exploiting sparsity of residuals in dynamic MRI & Dynamic MRI (cardiac imaging) \\ \hline
     Yun et al. \cite{yun2009high} & Exploiting a reference image for basis functions generation used to improve dynamic MRI & Dynamic MRI  (brain fMRI) \\ \hline
     Samsonov et al. \cite{samsonov_ismrm} & Exploiting sparsity of gradient of difference between baseline and follow-up scans & Longitudinal MRI \\ \hline
      Chen et al. \cite{chen2011framework} & Exploring the exploitation of a reference frame in x-t and x-f domains in dynamic MRI   & Dynamic MRI (cardiac imaging)\\ \hline
     Wu et al. \cite{wu2011prior} & Using noisy reconstruction as a reference for sorting in parallel imaging & Single-contrast MRI  \\ \hline
     Peng et al. \cite{peng2011reference} &  Exploiting reference image for sparsifying transform generation  & Single-contrast MRI \\ \hline 
     Bilgic et al. \cite{bilgic2011multi} & Exploit similarity of spatial derivatives in multi-contrast MRI & Multi-contrast MRI  \\ \hline
     Du et al. \cite{du2012compressed} and Lam et al. \cite{lam2011motion} & Exploiting similarity to a reference image in a CS-based hybrid reconstruction and registration scheme & Single-contrast MRI  \\ \hline
    Nguyen et al. \cite{nguyen2013modified} & Exploiting a reference image for generation of basis functions used for generalized series reconstruction of dynamic MRI & Dynamic MRI (brain fMRI)\\ \hline
	Haldar et al. \cite{haldar2013improved} & Using structural MRI for SNR improvement of DWI in an ML scheme & Diffusion MRI\\ \hline 
 Qu et al. \cite{qu2013magnetic,qu2014magnetic} & Exploiting similarity of image patches within and between multi-contrast MRI in CS framework & Multi-contrast MRI \\ \hline
	Huang et al. \cite{huang2014fast} & Joint TV and group wavelet based reconstruction for multi-contrast MRI & Multi-contrast MRI \\ \hline
 Chiew et al. \cite{chiew2014k} & Low-rank based reconstruction & Dynamic MRI (brain fMRI) \\ \hline
	  Li et al. \cite{li2015incorporation} & Using non reference-based reconstruction as a prior for reference-based reconstruction & Longitudinal MRI \\ \hline
	  Adluru et al. \cite{adluru2015mri} & Exploiting TV-based reconstruction for improved low-rank based reconstruction & Dynamic MRI (cardiac imaging) \\ \hline
	  Otazo et al. \cite{otazo2015low} & Low-rank based reconstruction & Dynamic MRI (cardiac imaging, MR angiography) \\ \hline   
       \hline 
      Our Method (FASTMER) &  Exploiting reference image in an adaptive-weighted CS scheme &  Single- and Multi-contrast MRI, Longitudinal MRI  \\ \hline   
       \end{tabular}  
       \label{table1}               
    \end{table*}
\end{center}

In this paper we present a scheme to exploit a reference image that is applicable for various MRI applications. The major contributions of this paper are: (a) exploiting similarity of intensity values to a reference scan is performed in an adaptive and weighted fashion, taking into account that the reference may exhibit major grey-level differences with respect to the current scan and; (b) several MRI imaging applications are tested (single-contrast high resolution MRI, multi-contrast MRI, and longitudinal MRI), with sampling performed via 2D radial sampling.

More specifically, we introduce a framework, called "FAST MRI by Exploiting a Reference scan" (FASTMER) which is based on two main elements: (a) exploiting both sparsity in the wavelet domain and similarity to a reference scan and (b) weighted reconstruction by adaptive selection of weights during the reconstruction process, taking into account the degree of similarity to the reference image. 
The adaptive weights in our framework adapt the reconstruction to the actual similarity between the scans. Therefore, it fits a variety of clinical imaging applications with supplemental imaging information, that in many cases is neglected due to its low fidelity. 

Experimental results demonstrate the applicability of the proposed method in three different MRI applications that utilize similarity to a reference image. The first application exploits similarity between two different imaging contrasts for fast scanning of one of them. The second example exploits similarity between different scans of the same patient for fast scanning of follow-up scans, and the third application exploits similarity between adjacent slices to improve SNR within the same imaging contrast.

The paper is organized as follows. Section \ref{sec:method} presents the proposed reference-based MRI approach. Section \ref{sec:results} describes experimental results. Section \ref{sec:discussion} discusses the method and Section \ref{sec:conclusions} concludes by highlighting the key findings of
the research.

\section{Reference-based MRI}
\label{sec:method}
\subsection{Compressed Sensing MRI}
The application of CS for MRI \cite{lustig2007sparse} exploits the fact that MRI scans are typically sparse in a transform domain, which is incoherent with the sampling domain. Nonlinear reconstruction is then used to enforce both sparsity of the image representation in some transform domain and consistency with the acquired data. A typical formulation of CS MRI recovery aims to solve the following unconstrained optimization problem (in a so-called Lagrangian form):
\begin{equation}
\begin{aligned}
& \underset{\vec{x}}{\text{min}}
&  \|\mat{F}_u\vec{x}-\vec{y}\|_2^2+\lambda\|\mat{\Psi}\vec{x}\|_1,
\end{aligned}
\label{eq05}
\end{equation}
where $\vec{x}\in \mathbb{C}^{N}$ is the $N$-pixel complex image to be reconstructed, represented as a vector, $\vec{y}\in \mathbb{C}^{M}$ denotes the k-space measurements, $\mat{F}_u$  is the undersampled Fourier transform operator, $\mat{\Psi}$ is a sparsifying transform 
operator and $\lambda$ is a properly chosen regularization parameter. We focus on brain MRI, known to be sparse in the wavelet domain. Therefore, we will assume throughout that $\mat{\Psi}$ is an appropriately chosen wavelet transform. 

This fundamental CS MRI formulation is the basis for many MRI reconstruction applications, where the sparse transform domain varies depending on the particular setting \cite{lang2008accelerating,jung2009k,gamper2008compressed,lustig2006kt}. We note that this basic formulation does not take into account any image-based prior information, that exists in many MRI applications.

\subsection{Reference-based compressed sensing MRI}
In many MRI imaging scenarios, an a-priori image that may exhibit similarity to the acquired image, is available. This image is coined hereinafter the ``reference image" is represented as  $\vec{x}_0$. A reference image could be a different imaging contrast in the same scan, an adjacent image slice or a previous scan of the same patient.  

In some imaging applications, we can assume that $\vec{x}_0$ and $\vec{x}$ are similar in most image regions \cite{gamper2008compressed}. Therefore the difference $\vec{x}-\vec{x}_0$ can be modelled as sparse, and a CS based optimization may utilize the reference image for improved reconstruction, via $\ell_1$ minimization. Such reference-based CS takes into account the fidelity of the measurements and the similarity to the reference scan, as follows:   
\begin{equation}
\begin{aligned}
& \underset{\vec{x}}{\text{min}}
&  \|\mat{F}_u\vec{x}-\vec{y}\|_2^2+\lambda{\|\vec{x}-\vec{x}_0\|_1}.
\end{aligned}
\label{eq1}
\end{equation}
\noindent This optimization problem assumes high degree of similarity between $\vec{x}_0$ and $\vec{x}$, and is therefore suitable for some specific MRI applications, such as dynamic MRI. However, many MRI applications do not utilize available reference imaging information (for instance, by using (\ref{eq1})) due to the fact that the similarity to the acquired image is partial, not guaranteed or unknown. 

We introduce a framework for reference based MRI, which takes into account the fact that $\vec{x}_0$ may exhibit differences versus $\vec{x}$. We also account for the fact that the vector $\vec{y}$ may represent multiple images that are contaminated with noise at different levels; we want to prioritize images with low noise standard deviation over ones with high noise standard deviation in the reconstruction process. Our approach is based on enforcing similarity between $\vec{x}$ and $\vec{x}_0$ via a weighted $\ell_1$ norm: 
\begin{equation}
\begin{aligned}
\underset{\vec{x}}{\text{min}}
\|\mat{A}(\mat{F}_u\vec{x}-\vec{y})\|_2^2+\lambda_1{\|\mat{W}_1\mat{\Psi}\vec{x}\|_1}+\lambda_2{\|\mat{W}_2(\vec{x}-\vec{x}_0)\|_1}
\end{aligned}
\label{eq2}
\end{equation}
\noindent where $\mat{A}$ is a diagonal matrix that controls the weight given to the fidelity of certain measurements (used to prioritize samples taken from images with low noise standard deviation). The matrices $\mat{W}_1$ and $\mat{W}_2$ are weighting matrices, $\mat{W}_k=\text{diag}([w_k^1,w_k^2,...,w_k^N])$ with $0\leq w_k^i \leq 1$, that control the weight given to each element in the sparse representation. In particular, $\mat{W}_1$  is used to weight specific wavelet atoms in the reconstruction process and $\mat{W}_2$ is used to weight image regions according to their similarity level with the reference scan. The parameters $\lambda_1$ and $\lambda_2$ are regularization parameters that control the weight given to each term in the optimization problem. 

In most cases, the expected noise level of the acquired data is known and the matrix $\mat{A}$ can be determined in advance. As to $\mat{W}_1$ and $\mat{W}_2$, there are cases in which neither the similarity to the reference image nor the support in the wavelet domain are known in advance, and therefore these weighting matrices have to be determined during the acquisition process as we describe in the next section.

\subsection{Adaptive weighting for reference based MRI}
\label{adaptive_rec}
Since the similarity of $\vec{x}$ to $\vec{x}_0$, as well as the support of $\vec{x}$ in the wavelet domain, are unknown, we estimate the matrices $\mat{W}_1$ and $\mat{W}_2$ from the acquired data, in an adaptive fashion. 
Inspired by Weighted-CS \cite{candes2008enhancing}, we propose an iterative reconstruction algorithm, where in each iteration a few k-space samples are added to the reconstruction process, based on their distance from the origin of the k-space (samples closer to the origin of the k-space are added first). The rationale lies in the the structure of the minimization problem (\ref{eq2}). Due to the fact that (\ref{eq2}) is a non-convex minimization problem, we would like to avoid convergence to local minima. Therefore, we experimentally experienced that when we gradually add k-space samples at each iteration the convergence process is more dynamic and improved results are obtained. At the end of each iteration, $\vec{\hat{x}}$ is estimated, to serve as the basis for estimating the weighting matrices in the next iteration. 

Our rationale behind the iterative computation of $\mat{W}_k$ is as follows. For $\mat{W}_1$, we would like to relax the demand for sparsity on elements in the support of $\mat{\Psi}\vec{x}$. For $\mat{W}_2$, we would like to enforce sparsity only in spatial regions where $\vec{x}\approx\vec{x}_0$.  We note that when one of the weights is given a small value, for instance, $w_2^i \rightarrow 0  $, then the sparsity on the corresponding image pixel, $\mathbf{x}_i$, is relaxed, and vice-versa; when $w_2^i \rightarrow 1  $, sparsity is enforced on $\mathbf{x}_i$ (i.e. the $\ell_1$ minimization would prefer solutions where $\mathbf{x}_i \rightarrow 0$).

Since $\vec{x}$ is unknown, $\vec{\hat{x}}$, updated in every iteration, is used instead. 
The elements of the weighting matrices are then chosen as follows:
\begin{align}
\begin{split}
 w_1^i&=\frac{1}{1+[|\mat{\Psi}\vec{\hat{x}}|]_i} \\ 
 w_2^i&=\frac{1}{1+[|\vec{\hat{x}}-\vec{x}_0|]_i}
\label{eq6}
\end{split}
\end{align}

\noindent where $[\cdot]_i$ denotes the $i$th element of the vector in brackets. We note that in a similar way to Cand{\`e}s' approach \cite{candes2008enhancing}, developed for $\vec{x}$ which is truly sparse, the weights in (\ref{eq6}) are given values that vary between $0$ and $1$. The values for $w_1^i$ and $w_2^i$ are inversely proportional to those of the corresponding elements in the vectors $\mat{\Psi x}$ and $\vec{\hat{x}}-\vec{\hat{x}}_0$, respectively.  Therefore, since reconstruction quality improves at each consecutive iteration, we get $w_2^i \rightarrow 1$ in regions where $\mathbf{x} \sim \mathbf{x}_0$, thereby enforcing sparsity of the difference $\mathbf{x}-\mathbf{x}_0$ in those regions. The same analysis applies for $\mathbf{W}_1$ and the representation of the image in the wavelet domain.

The proposed algorithm is coined FAST MRI by Exploiting a Reference scan (FASTMER) and is summarized in Algorithm 1. Note that in the first iteration of the algorithm we do not assume similarity with the reference image (i.e., we set $\mat{W}_1=\mat{I}$ and $\mat{W}_2=\mat{0}$ for the first iteration). This is done in order to prevent the algorithm from convergence to an incorrect solution in cases where  similarity between scans does not exist.  

\newlength\myindent
\setlength\myindent{10em}
\newcommand\bindent{%
  \begingroup
  \setlength{\itemindent}{\myindent}
  \addtolength{\algorithmicindent}{\myindent}
}
\newcommand\eindent{\endgroup}
\renewcommand{\algorithmicrequire}{\textbf{Input:}}
\renewcommand{\algorithmicensure}{\textbf{Output:}}
\begin{algorithm}[H]
\caption{Fast MRI by Exploiting Reference (FASTMER)}
\label{algo1}
 \begin{algorithmic} 
 \renewcommand{\algorithmicrequire}{\textbf{Input:}}
\renewcommand{\algorithmicensure}{\textbf{Output:}}
\REQUIRE \hspace{3mm} \\
Number of iterations: $N_I$; Reference image: $\vec{x}_0$ \\
Sampled k-space: $\vec{z}$; Tuning constants: $\lambda_1,\lambda_2$\\
Number of k-space samples added at each iteration: $N_k$\\
Expected fidelity of measurements: $\mat{A}$\\ 

\ENSURE Estimated image: $\mat{\hat{x}}$
 \renewcommand{\algorithmicrequire}{\textbf{Initialize:}}
 \REQUIRE \hspace{3mm} \\
 $\mat{W}_1=\mat{I}$, $\mat{W}_2=\mat{0}$; 
\renewcommand{\algorithmicrequire}{\textbf{Reconstruction:}}
 \REQUIRE \hspace{1mm} \\
\FOR{$l=1$ to $N_I$}
\STATE Add $N_k$ new samples to $\vec{y}$ from $\vec{z}$ according to distance from center of k-space.
\STATE {\bf Weighted reconstruction:} Estimate $\vec{\hat{x}}$ by solving (\ref{eq2})
\STATE {\bf Update weights: } Update $\mat{W}_1$ and $\mat{W}_2$ according to (\ref{eq6})\\
 \ENDFOR
\end{algorithmic}
\end{algorithm}

To solve the $\ell_1$-minimization problem (\ref{eq2}) in the weighted reconstruction phase, we use an extension of SFISTA \cite{tan2014smoothing}. The extended algorithm is summarized in Algorithm 2, where the notation $\|\cdot\|_2$ for matrices denotes the largest singular value.  The operator $\Gamma_{\lambda \mu }(\vec{z})$ is the soft shrinkage operator, which operates element-wise on $\vec{z}$ and is defined as (for complex valued $z_i$):

\begin{equation}
 \Gamma_{\lambda \mu }(z_i)=\begin{cases}
    \frac{|z_i|-\lambda\mu}{|z_i|}z_i, & |z_i|>\lambda\mu  \\
   0, & \text{otherwise}.
  \end{cases}
\label{eqA7}
\end{equation}

\begin{algorithm}[H]\caption{SFISTA algorithm for FASTMER}
\label{algo2}
\begin{algorithmic}
\REQUIRE \hspace{3mm} \\
k-space measurements: $\vec{y}$ \\
Sparsifying transform operator: $\mat{\Psi}$\\
An $N\times N$ k-space undersampling operator: $\mat{F}_u$\\
Reference image: $\vec{x}_0$\\
Expected fidelity of measurements: $\mat{A}$\\ 
Tuning constants: $\lambda_1,\lambda_2,\mu$\\
An upper bound: $L\geq \|\mat{AF}_u\|_2^2+\frac{\|\mat{W_1\Psi}\|_2^2+\|\mat{W_2}\|_2^2}{\mu}$
\ENSURE Estimated image: $\mat{\hat{x}}$
\renewcommand{\algorithmicrequire}{\textbf{Initialize:}}
 \REQUIRE \hspace{3mm} \\
\STATE $\vec{x}_1=\vec{z}_2=\mat{F}_u^*\vec{y}$, $t_2=1$
\renewcommand{\algorithmicrequire}{\textbf{Iterations:}}
 \REQUIRE \hspace{3mm} \\
\STATE {\bf Step k:} $(k\geq 2)$ Compute
\STATE $\nabla f(\vec{z}_k)= \mat{A}^*(\mat{F}_u^*(\mat{A}(\mat{F}_u\vec{z}_k-\vec{y})))$
\STATE $\nabla g_{1\mu}(\mat{W}_1\mat{\Psi}\vec{x}_{k-1})=\frac{1}{\mu}\mat{W}_1\mat{\Psi^*}(\mat{W}_1\mat{\Psi}\vec{x}_{k-1}-\Gamma_{\lambda_1\mu}\left(\mat{W}_1\mat{\Psi}\vec{x}_{k-1})\right)$
\STATE $\nabla g_{2\mu}(\mat{W}_2(\vec{x}_{k-1}-\vec{x}_0))=\frac{1}{\mu}\mat{W}_2(\mat{W}_2(\vec{x}_{k-1}-\vec{x}_0)-\Gamma_{\lambda_2\mu}\left(\mat{W}_2(\vec{x}_{k-1}-\vec{x}_0)\right)$
\STATE $\vec{x}_k=\vec{z}_k-\frac{1}{L}(\nabla f(\vec{z}_k)+\nabla g_{1\mu}(\mat{W}_1\mat{\Psi}\vec{x}_{k-1})+\nabla g_{2\mu}(\mat{W}_2(\vec{x}_{k-1}-\vec{x}_0)))$
\STATE $t_{k+1}=\frac{1+\sqrt{1+4t_k^2}}{2}$
\STATE $\vec{z}_{k+1}=\vec{x}_k+\frac{t_k-1}{t_{k+1}}(\vec{x}_k-\vec{x}_{k-1})$
\end{algorithmic}
\end{algorithm}

Algorithm 2 minimizes  (\ref{eq2}), where the trade-off between the two sparsity assumptions is controlled by the ratio between $\lambda_1$ and $\lambda_2$, via $\Gamma(\cdot)$, and the overall convergence is controlled by $\mu$.

\section{FASTMER for SNR improvement}
\label{extention_snr}
In MRI, SNR is proportional to the number of protons involved in generating the measured signal. As a result, thick slices provide better SNR than thin ones. However, the thinner the slice, the better the image resolution in the z-axis. Therefore, to obtain high quality MRI for clinical evaluation purposes, high SNR MRI that consists of thin slices is required. The common approach today for SNR improvement of MRI with thin slices consists of averaging over several excitations (usually three or four), which extends the scanning time by the same amount.

In this application, where thin slices are acquired, one may consider shortening scanning time by reducing the number of excitations and exploiting similarity between thin slices. In addition, similarity to a thick, high SNR image slice that overlaps the thin slices can also be utilized for SNR improvement.

We will consider a specific implementation, where a single excitation is used to acquire two thin adjacent slices with low SNR, $\vec{x}_1$ and $\vec{x}_2$, and a single thick slice, $\vec{x}_3$ that spatially overlaps $\vec{x}_1$ and $\vec{x}_2$. Obviously, our goal is to improve the SNR of $\vec{x}_1$ and $\vec{x}_2$ taking into account the similarity between them and the high-SNR, thick slice, $\vec{x}_3$. In this specific example we perform a total of three acquisitions for high SNR reconstruction of two adjacent slices, instead of 8 acquisitions (4 excitations for each thin slice) in the conventional, multiple excitations-based approach.

To introduce the method within our framework, let 
\begin{equation}
 \vec{y}=\begin{bmatrix}
    \vec{y}_1\\
    \vec{y}_2\\
    \vec{y}_3
  \end{bmatrix}
,\ \ \vec{x}= \begin{bmatrix}
    \vec{x}_1\\
    \vec{x}_2\\
    0.5(\vec{x}_1+\vec{x}_2)
  \end{bmatrix}
\end{equation}
\noindent where $\vec{y}$ represents the k-spaces of two thin slices and the corresponding thick one, respectively, and $\vec{x}$ represents the two thin slices and their average, which corresponds to the overlapping thin slice. The matrix $\mat{A}$ is determined by the estimated noise level of the elements in $\vec{y}$, such that
$\mat{A}=\textrm{diag}(\frac{1}{\sigma_1}\mat{I}_N, \frac{1}{\sigma_2}\mat{I}_N, \frac{1}{\sigma_2}\mat{I}_N )$ where $\{\sigma_i\}_{i=1}^{3}$ are the noise standard deviation of $\{\vec{y}_i\}_{i=1}^{3}$, respectively and $\mat{I}_N$ is an identity matrix of size $N$.
Similarity is enforced between the thin slices, and  (\ref{eq2}) is reformulated as:
\begin{equation}
\begin{aligned}
\underset{\vec{x}}{\text{min}}
\|\mat{A}(\mat{F}_3\vec{x}-\vec{y})\|_2^2+\lambda_1{\|\mat{W}_1\mat{\Psi}_3\vec{x}\|_1}+\lambda_2{\|\mat{W}_2\mat{B}\vec{x}\|_1.}
\end{aligned}
\label{eq12}
\end{equation} 
\noindent Here $\mat{F}_3=\textrm{diag}([\mat{F},\mat{F},\mat{F}])$ is a block diagonal matrix, with three Fourier matrices on the main diagonal , $\mat{\Psi}_3=\textrm{diag}([\mat{\Psi},\mat{\Psi},\mat{\Psi}])$ and $\mat{B}=[\mat{I}_N \quad -\mat{I}_N \quad \mat{0}]$.


Similarity between the thick slice to the average between the thin slices is enforced in the Fourier domain, via the leftmost term of (\ref{eq12}). Adapting Algorithm 1 and Algorithm 2 to solve (\ref{eq12}) is straightforward, the final method appears in Appendix A.

%

\section{Experimental Results}
\label{sec:results}
To demonstrate the performance of our reference based MRI approach we examine three MRI applications, all of which utilize a reference scan for improved reconstruction. Where relevant, partial k-space acquisition was obtained by 2D radial sampling of a fully sampled k-space. The angles of the trajectories were taken randomly according to a uniform distribution. Non-uniform sampling and reconstruction was performed using the non-uniform Fourier transform (NUFFT) package of Fessler et al. \cite{fessler2003nonuniform}. A Daubechies-4 wavelet transform was used as the sparsifying transform. Different values of $\lambda_1,\lambda_2$ in the range of [0, 0.9] were examined, and the best result in terms of image quality is presented in each case (see Section \ref{sensitivity_analysis}). We used $\mu=10^{-3}(\frac{\lambda_1+\lambda_2}{2})^{-1}$ in our experiments. The number of iterations used in Algorithm 2 for the results obtained in this section is between 30 and 50 iterations.

All scans were performed on a GE Signa 1.5T HDx scanner, using a 8-channel head coil and with matrix size of $320 \times 320$ and FOV of $20cm$ for each in-plane direction. High SNR images reconstructed from fully sampled, multiple excitations data serve as the gold standard. The source code and data required to reproduce the results presented in the this paper can be downloaded from: \url{http://www.technion.ac.il/~weizmanl/software}

To provide a quantitative measure for the results, we examine the peak signal-to-noise ratio (PSNR) of each experiment, defined as: $ \textrm{PSNR}=10\text{log}_{10}({M^2/{V_s})}$, where $M$ denotes the maximum possible pixel value in the image and $V_s$ is the Mean Squared Error (MSE) between the original image, $\vec{x}$ and the reconstructed image, $\hat{\vec{x}}$. The resulting weighting matrices ($\mat{W}_1$ and $\mat{W}_2$) are presented in Appendix \ref{AppB}, as well as the similarity maps between the reference image and the reconstructed image.

Since the proposed approach is based on the difference between the reference image from the reconstruction, we need to verify that both images are aligned and have matched intensities. In the experiments presented hereinafter we conducted a wavelet-based reconstruction first (according to (\ref{eq05})) using the acquired data, to get a rough estimate on the alignment parameters and the grey-level range of the reconstructed image. Then, we embedded the extracted parameters in our reconstruction process for successful reference-based reconstruction. This issue is further discussed in Section \ref{practical_limitations}.

\begin{figure*}
\centering 
\includegraphics[width=90pt,trim=1.9cm
1cm 2cm 1.1cm,
clip=true]{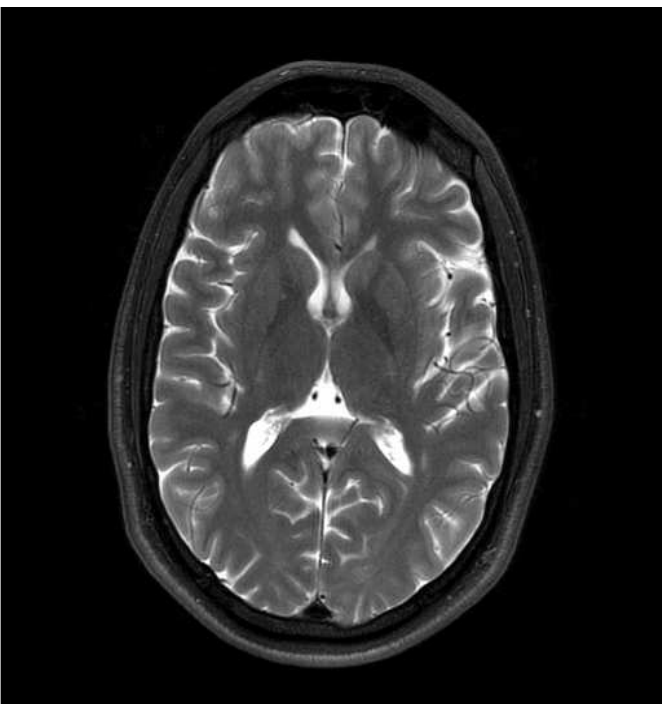}
\includegraphics[width=90pt,trim=1.9cm
1cm 2cm 1.1cm,
clip=true]{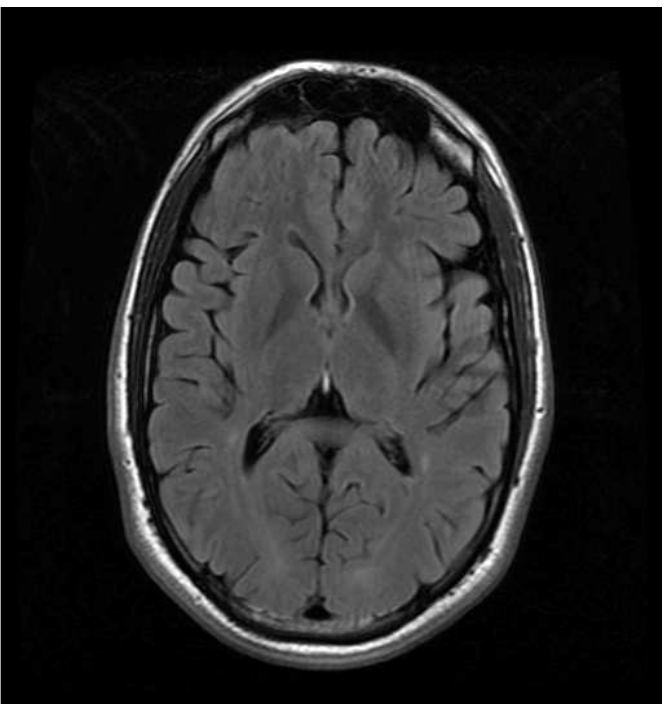}
\includegraphics[width=90pt,trim=1.9cm
1cm 2cm 1.1cm,
clip=true]{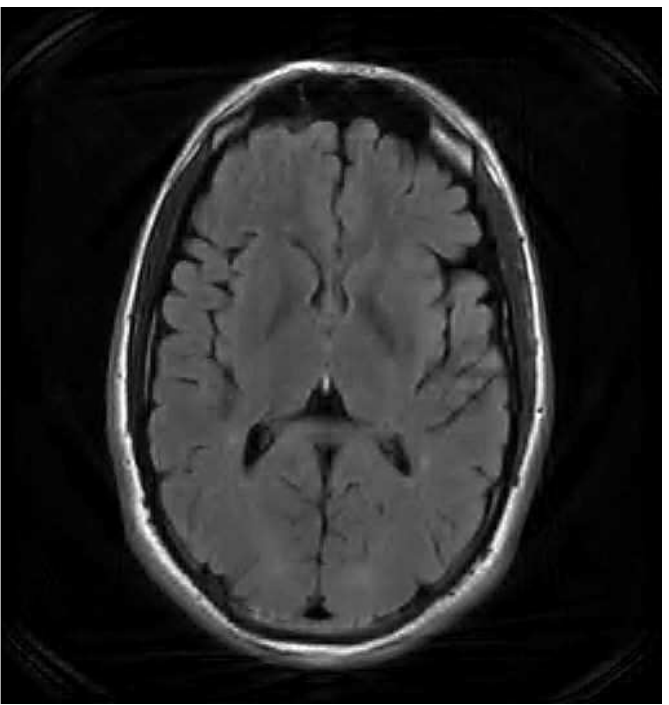}
\includegraphics[width=90pt,trim=1.9cm
1cm 2cm 1.1cm,
clip=true]{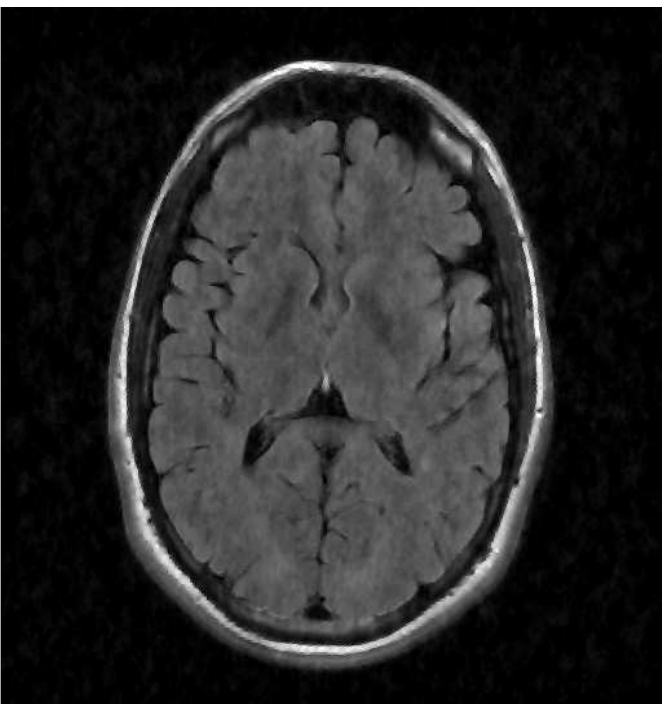}
\includegraphics[width=90pt,trim=1.9cm
1cm 2cm 1.1cm,
clip=true]{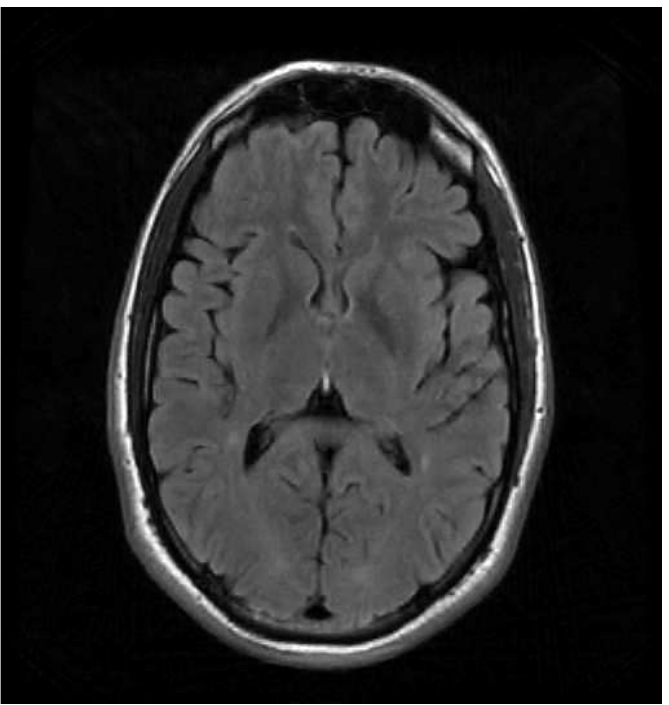}\\
\vspace{-3mm}{\footnotesize \begin{flushleft}\hspace{15mm} T2-weighted \hspace{17mm} FLAIR \hspace{15mm}  Wavelet+TV result\hspace{5mm} Bilgic et al. \cite{bilgic2011multi} result\hspace{7mm} FASTMER result\\ \vspace{0mm}
\hspace{14mm}(gold standard) \hspace{9mm} (gold standard) \hspace{12mm} (PSNR=28dB) \hspace{13mm}(PSNR=30dB) \hspace{10mm} (PSNR=33dB)\end{flushleft}}

\caption{FASTMER used within the same scan: reconstruction results utilizing similarity between T2 and FLAIR contrasts. The two leftmost images are the T2 and FLAIR images reconstructed from 100\% of k-space data. The three rightmost images show reconstruction results from 25\% of k-space FLAIR data, for wavelet+TV based reconstruction, the method of Bilgic et al. for multi-contrast reconstruction using Bayesian-CS \cite{bilgic2011multi} and our proposed, reference-based approach (FASTMER). The numbers in brackets are the PSNR values vs. the FLAIR gold standard.  
}
\label{fig2}

\end{figure*}

\subsection{Utilizing similarity between T2-weighted and FLAIR}
\label{t2_flair_experiment}
In this experiment, our goal is to reconstruct a FLAIR image, $\vec{x}$, from undersampled measurements, utilizing similarity to a T2-weighted image. Images were acquired with slice thickness of $4mm$. FLAIR acquisitions parameters were: TE=$123ms$,TR=$8000ms$ and TI=$2000ms$, and T2-weighted acquisition parameters were: TE=$68.4ms$ and TR=$6880ms$. We sampled only 25\% of the FLAIR k-space with radial sampling and utilized the fully sampled T2-weighted scan as the reference image $\vec{x}_0$. Since all samples were acquired with similar noise level, $\mat{A}=\mat{I}$.

\begin{figure*}
\vspace{5mm}
\centering 
\includegraphics[width=112pt,trim=2cm
4.5cm 6cm 4cm,
clip=true]{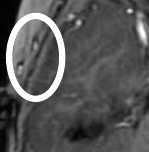}
\includegraphics[width=112pt,trim=4cm
4.8cm 6.45cm 5.4cm,
clip=true]{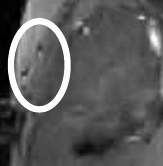}
\includegraphics[width=112pt,trim=4cm
4.8cm 6.45cm 5.4cm,
clip=true]{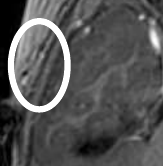}
\includegraphics[width=112pt,trim=4cm
4.8cm 6.45cm 5.4cm,
clip=true]{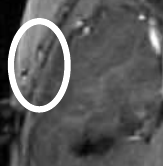}\\
\vspace{-3mm}{\footnotesize \begin{flushleft}\hspace{15mm} Gold standard (zoom) \hspace{7mm} Wavelet+TV recon. (zoom) \hspace{6mm} Samsonov et al. (zoom) \hspace{7mm}  FASTMER (zoom)\\

\hspace{14mm}\hspace{45mm} (PSNR=34dB) \hspace{20mm}(PSNR=35dB) \hspace{18mm}  (PSNR=37dB)\end{flushleft}}
\begin{minipage}[c]{0.3\textwidth} \vspace{-60mm}
\includegraphics[width=240pt,trim=0.7cm
1cm 0.79cm 1.5cm,
clip=true]
{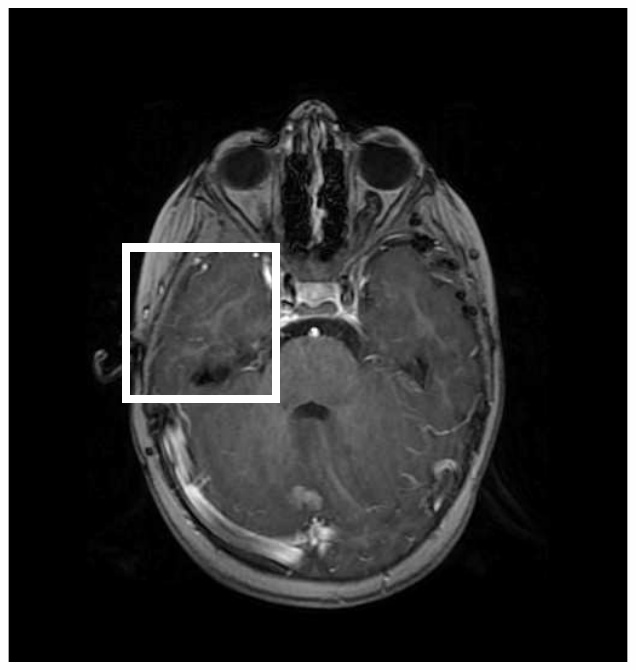}\\
\vspace{-5mm}{\footnotesize \begin{flushleft}\hspace{16mm} Gold standard \end{flushleft}}
\end{minipage}
  \begin{minipage}[c]{0.67\textwidth}
   \caption{FASTMER used in longitudinal studies: reconstruction results from 25\% of k-space data. Bottom left: ground truth follow-up image reconstructed from 100\% of k-space data. The rectangle defines the region-of-interest explored in this figure. Upper row: Enlarged versions of the region-of-interest in the ground truth image (leftmost image), followed by the results of wavelet+TV based reconstruction, the method of Samsonov et al. for longitudinal studies \cite{samsonov_ismrm} and our proposed, reference-based approach (FASTMER). The numbers in brackets are the PSNR values vs. the gold standard. It can be seen that FASTMER exhibits results which are very similar to the gold standard, and reveals imaging features that are blurred or not visible in other recoveries.}
\label{fig51}
  \end{minipage} 
\end{figure*}

To provide a baseline for comparison, we compared FASTMER to two methods. The first is reconstruction based on sparsity in the wavelet domain and enforces total-variation (wavelet+TV) \cite{lustig2007sparse} using undersampled FLAIR data only. The second is the algorithm of Bilgic et al. \cite{bilgic2011multi} which exploits the similarity of gradients between different contrasts. 

Figure \ref{fig2} shows the fully sampled T2 and FLAIR images, the wavelet+TV based reconstruction, the reconstruction using the method of Bilgic et al. \cite{bilgic2011multi} and the result of FASTMER. In addition, the PSNR values of each reconstruction method vs. the gold-standard are provided. It can clearly be seen that FLAIR reconstruction with FASTMER outperforms the two other methods used for comparison, using only 25\% of the data.

\subsection{Utilizing similarity between baseline and follow-up scans}

Repeated brain MRI scans of the same patient every few weeks or months are very common for follow-up of brain tumors. Here, our goal is to use a previous scan in the time series as a reference scan for reconstruction of a follow-up scan. In this application we need to take into account that similarity between the reference and current scans is not guaranteed (e.g. due to pathology changes), and prior information on spatial regions that may exhibit differences is not available. These obstacles are discussed thoroughly in our previous publication \cite{weizman2014application} and in Section \ref{practical_limitations} and  Appendix \ref{AppC} of this paper. 
Since all samples are acquired with similar noise level, we set $\mat{A}=\mat{I}$.

We compared FASTMER to two methods. The first is wavelet+TV based reconstruction. The second is the method of Samsonov et al. \cite{samsonov_ismrm} which exploits the gradient images similarity between a follow-up scan and a baseline scan using Bayesian-CS, in a non-weighted approach.

Figure \ref{fig51} shows reconstruction results of a follow-up contrast enhanced T1-weighted brain scan utilizing the baseline scan as reference (TE=$11.5ms$, TR=$520ms$ slice thickness: $1mm$ for both scans). Results were obtained using only 25\% of k-space data. It can be seen that FASTMER exhibits imaging features that are hardly visible in both of the methods it is compared against. In addition, the PSNR values of each reconstruction algorithm vs. the gold-standard are provided. The superiority of our approach is achieved thanks to the iterative mechanism that adapts the reconstruction to match actual similarity. 

\subsection{Utilizing similarity between adjacent slices} 
In this application we examine the extension of fast reference based MRI detailed in Section \ref{extention_snr} to improve SNR of thin MRI slices. We acquired a brain T2-weighted scan with slice thickness of $0.8mm$ followed by an additional acquisition with slice thickness of $1.6mm$ (TE=$68.4ms$ and TR=$6880ms$ for all scans). In all scans a single excitation was used. As a result, we obtained a low SNR scan consisting of thin slices, and high SNR scan consisting of thick slices where each thick slice overlaps two thin ones. Our goal is to reconstruct a high SNR scan comprised of thin slices from this data.

\begin{figure*}
\centering
\begin{turn}{90} 
\hspace{20mm} Slice I
\end{turn} 
\includegraphics[width=108pt,trim=1.3cm
0.8cm 1.5cm 1.5cm,
clip=true]{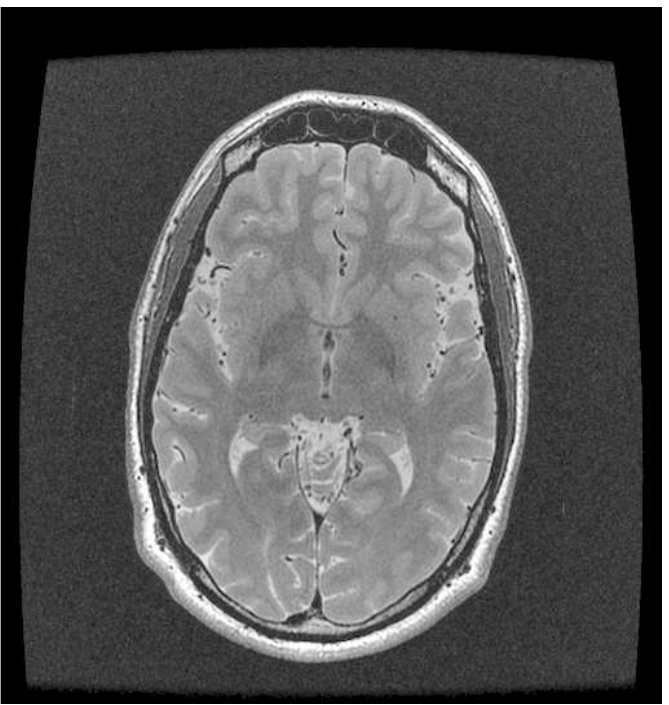}
\includegraphics[width=108pt,trim=1.3cm
0.8cm 1.5cm 1.5cm,
clip=true]{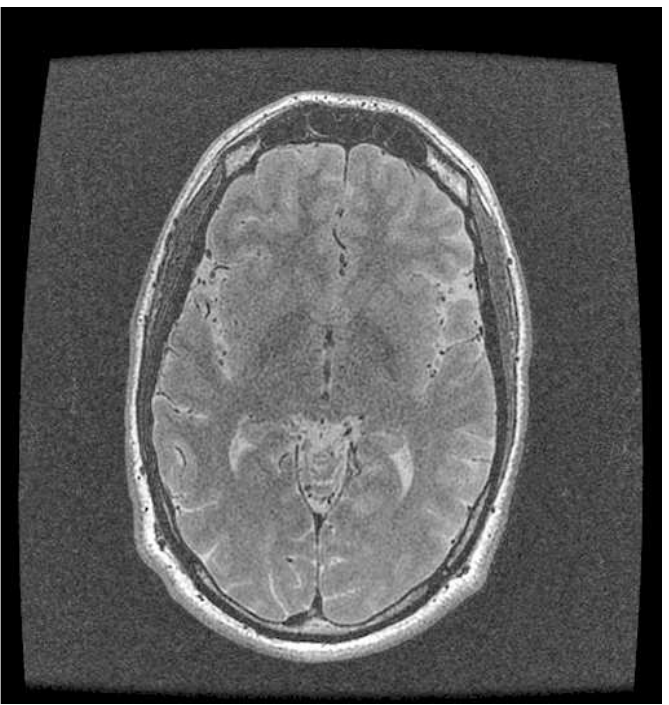}
\includegraphics[width=108pt,trim=1.3cm
0.8cm 1.5cm 1.5cm,
clip=true]{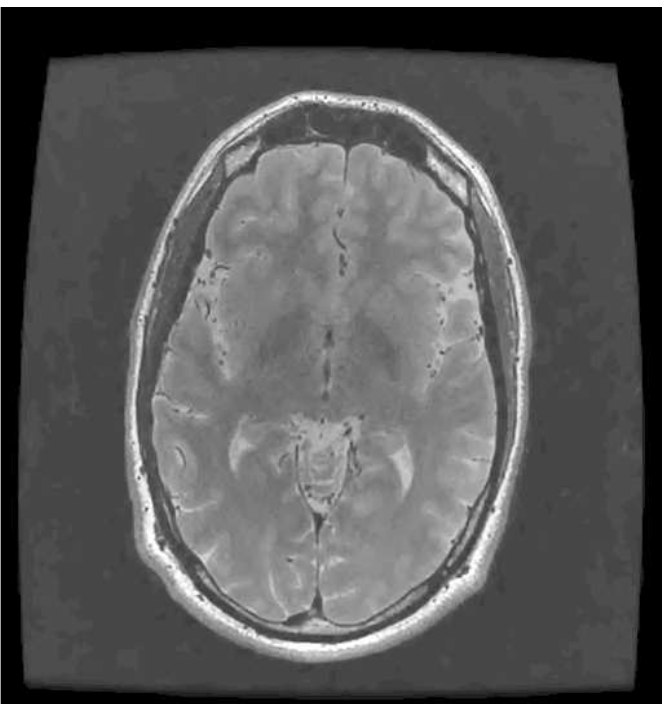}
\includegraphics[width=108pt,trim=1.35cm
0.8cm 1.5cm 1.5cm,
clip=true]{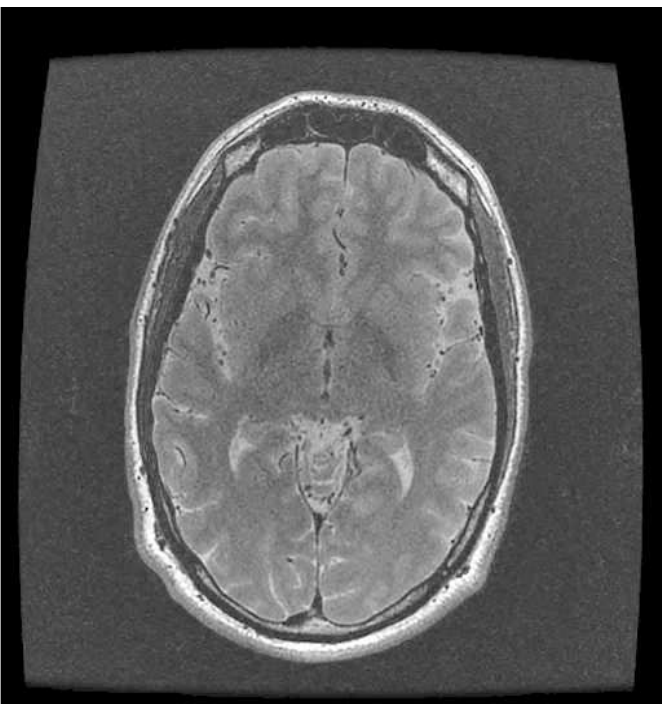}\\

\vspace{2mm}
\begin{turn}{90} 
\hspace{20mm}Slice II
\end{turn} 
\includegraphics[width=108pt,trim=1.3cm
0.8cm 1.5cm 1.5cm,
clip=true]{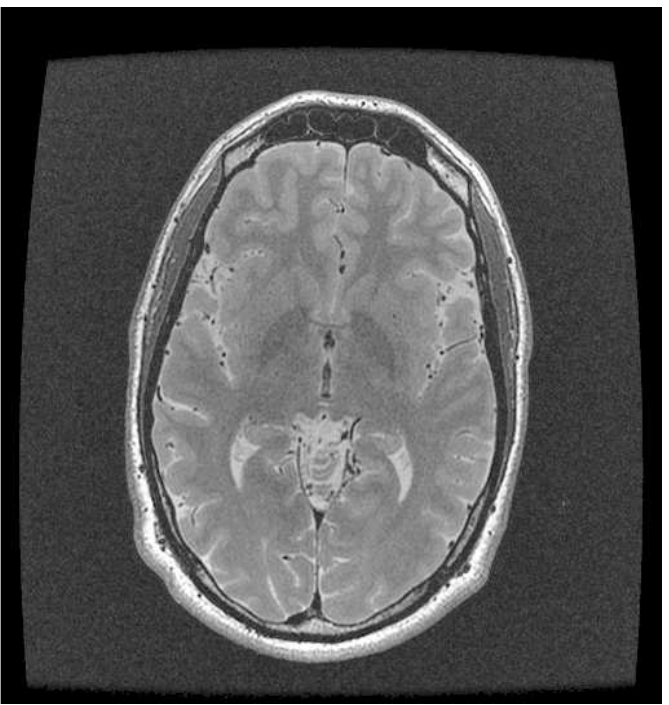}
\includegraphics[width=108pt,trim=1.3cm
0.8cm 1.5cm 1.5cm,
clip=true]{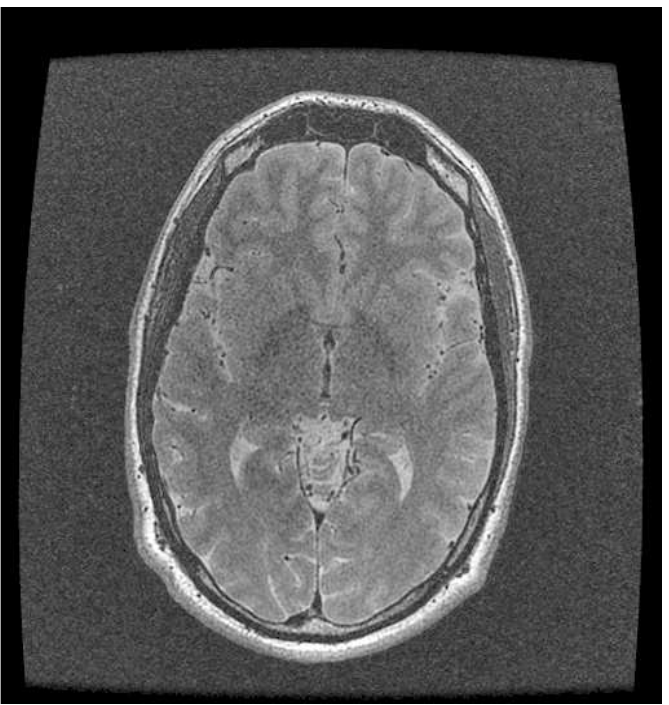}
\includegraphics[width=108pt,trim=1.3cm
0.8cm 1.5cm 1.5cm,
clip=true]{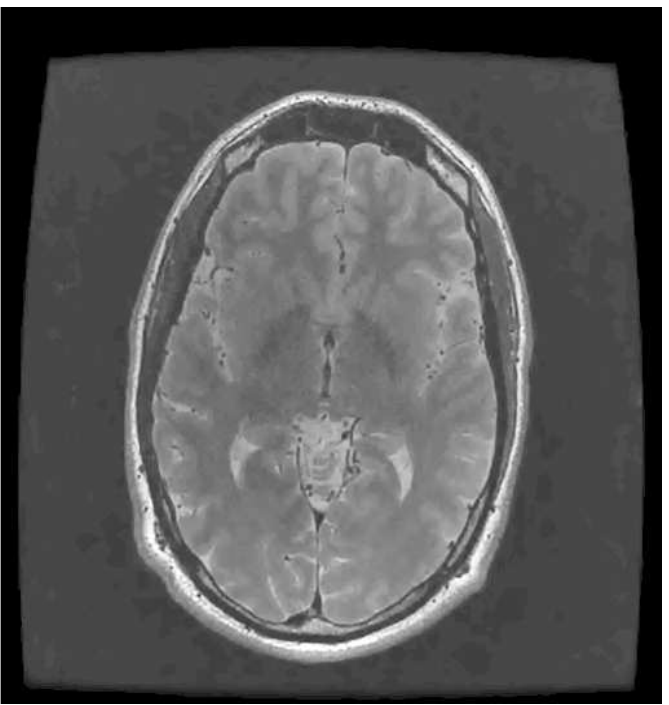}
\includegraphics[width=108pt,trim=1.35cm
0.8cm 1.5cm 1.5cm,
clip=true]{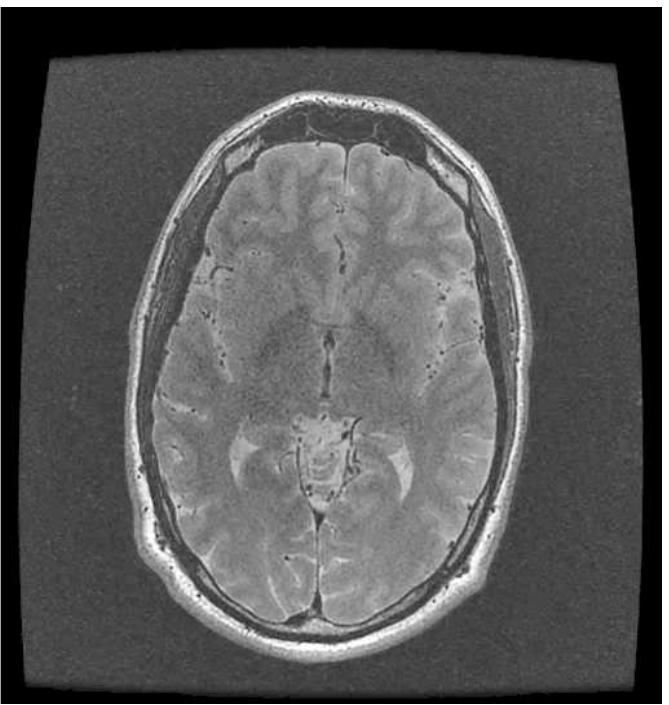}\\
\vspace{-3mm}{\footnotesize \begin{flushleft}\hspace{22mm} Gold standard \hspace{12mm} Original noisy images\hspace{12mm}  Wavelet+TV result\hspace{15mm} FASTMER result\\ \vspace{-1mm}
\hspace{25mm}(NEX=4) \hspace{24mm} (NEX=1) \hspace{20mm} (PSNR=31dB) \hspace{20mm}(PSNR=34dB) \end{flushleft}}

\caption{FASTMER used within the same imaging contrast: reconstruction results from low SNR data. Each row corresponds to a single slice, from two adjacent slices. It can be seen that high similarity exists within adjacent slices, which can be exploited to improve SNR. The left most column shows the gold standard, acquired using four excitations (NEX=4). The input to our approach is presented as the original noisy images, acquired with single acquisition (NEX=1), followed by the result of a Total-Variation (TV)-based reconstruction. The rightmost column shows FASTMER recovery. The numbers in brackets are the PSNR values vs. the gold standard (values were averaged over the two slices used in the experiment).}

\label{fig1}
\end{figure*}
\begin{figure*}
\centering
\includegraphics[width=135pt]{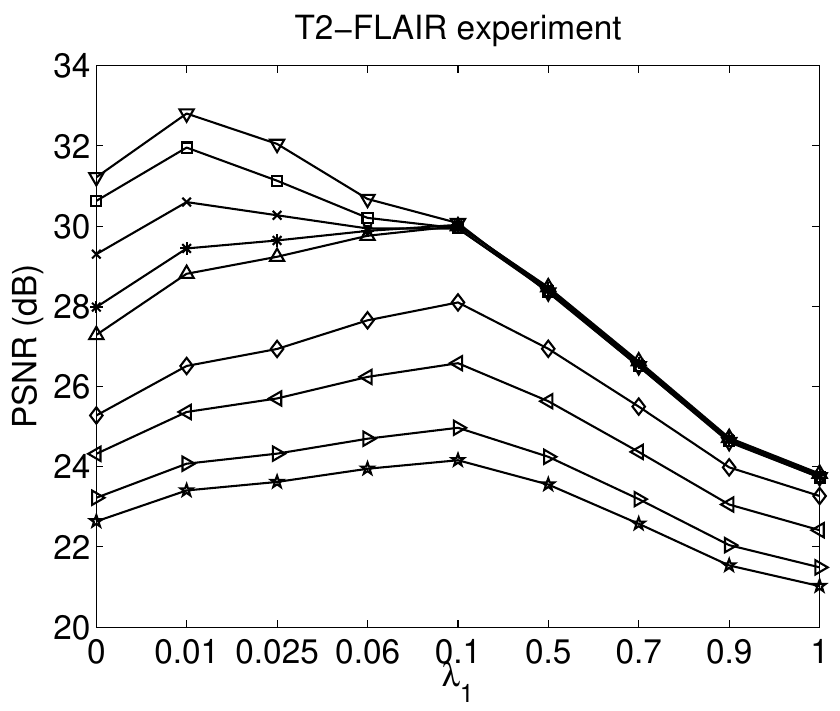}\hspace{2.5mm}
\includegraphics[width=135pt]{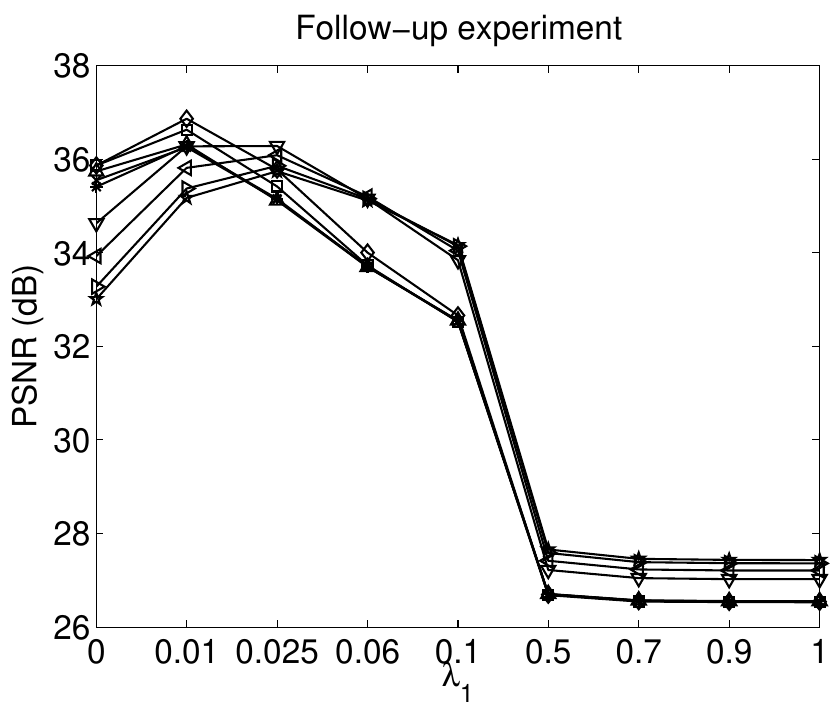}\hspace{2.5mm}
\includegraphics[width=135pt]{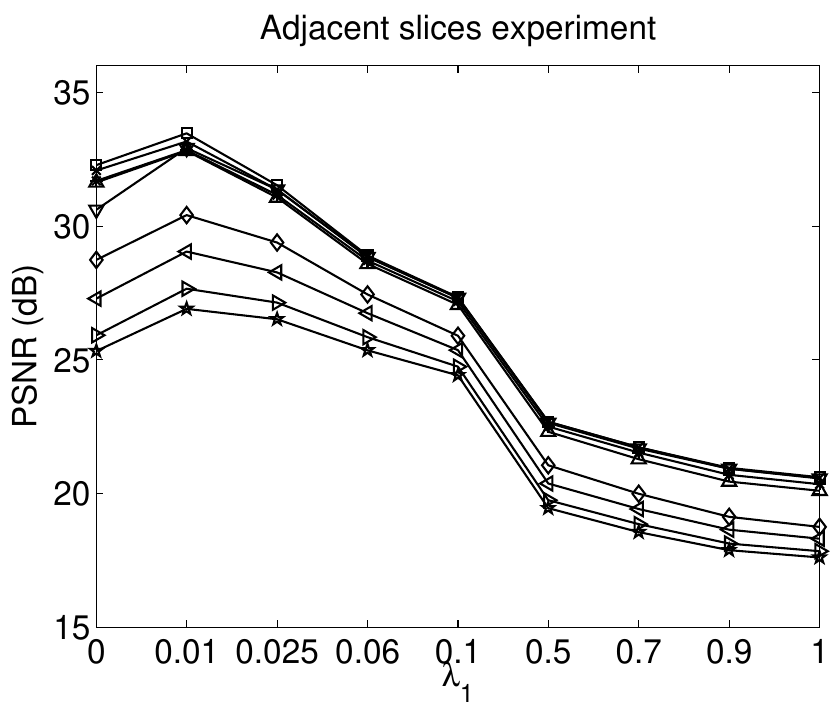}
\begin{minipage}[l]{0.05\textwidth}\vspace{-4.2cm} \includegraphics[width=40pt,trim=0cm 0cm 0cm 0cm,clip=true]{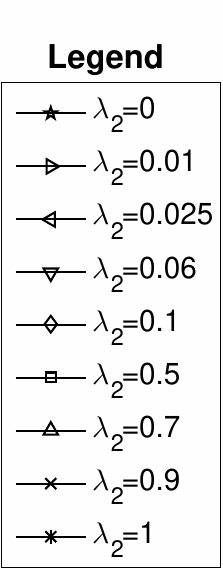}
\vspace{4.5cm}
\end{minipage}

\caption{Sensitivity analysis: PSNR results of T2-FLAIR (left), follow-up (middle) and adjacent slice similarity (top) applications for various values of $\lambda_1$ and $\lambda_2$. Each line represents a different value of $\lambda_2$, according to the legend at the left. }
\label{fig45}
\end{figure*}

Figure \ref{fig1} shows  thin slices acquired  using 4 excitations (NEX=4), used as our gold standard in this experiment, and the noisy input images that were acquired with a single excitation (NEX=1). Here, we compare our method to a wavelet+TV based approach, which has been tested previously to improve SNR in MRI \cite{ma2008efficient}.

In terms of scanning time, 4 excitations are required to obtain thin slices with SNR comparable to SNR of data reconstructed with FASTMER. Therefore, without additional acceleration techniques (parallel imaging etc.), our approach requires scanning 3 slices once versus scanning 2 slices 4 times in conventional scanning, yielding a speed-up factor of 2.6.

\subsection{Parameter sensitivity analysis}
\label{sensitivity_analysis}
In this analysis we examine the sensitivity of FASTMER to changes in the regularization parameters of the algorithm. Figure \ref{fig45} shows the PSNR results as $\lambda_1$ and $\lambda_2$ vary. The values shown for the adjacent slices experiment were averaged over the two thin slices used in the experiment. Generally, we observe that lower values of 
$\lambda_1$ and $\lambda_2$, with $\lambda_1<\lambda_2$ provide reasonable PSNR. This can be explained by the fact that over-promoting sparsity versus consistency to measurements degrades the reconstruction quality. In addition, we see that the T2-FLAIR experiment provides a lower range of PSNR values in comparison to other experiments. This can be explained by the fact that similarity is not enforced over the entire image in this case, due to many regions of differences between FLAIR and T2. 

\section{Discussion}
\label{sec:discussion}
\subsection{Theoretical justification}
The algorithm presented in our paper generalizes the reweighted $\ell_1$ minimization method by Candes, Wakin, and Boyd \cite{candes2008enhancing} to the case where there is side information (the reference-image in FASTMER). Their algorithm is quite intuitive and works very well in practice, and their analytical justification can easily be extended to our case. To analyze the theoretical aspects of our weighting algorithm,  we explored our model from a signal processing point of view \cite{mota2016reference}. 

We developed a bound on the number of measurements required for perfect reconstruction of $\vec{x}$ with high probability for the case where side information exists. We derived a weighted reconstruction scheme based on the bound. We have shown that adding the weights to our scheme highly improves the results in comparison to other non-weighted $\ell_1$-minimization based solutions. The results of this work have been published \cite{mota2016reference} and the reader is referred to this recent publication to delve into the theoretical aspects of this work.

\subsection{Adaptive sampling}
In our approach all data is acquired at once. Several recent publications suggest that prior knowledge can also be used to optimize data acquisition \cite{zientara1994dynamically,nagle1999multiple,gao2000optimal}. However, since in our framework the similarity between scans is not guaranteed, we avoid using prior information during sampling. Another way of utilizing the reference image in the sampling stage would be to acquire a small number of samples in each iteration based on the reconstruction results, in an adaptive manner \cite{haupt2009adaptive,haupt2011distilled,wei2013multistage,seeger2010optimization,ravishankar2011adaptive}. This approach, which requires image reconstruction at each iteration as part of the sampling process, has been tested in our previous work \cite{weizman2014application}. It was shown to be time consuming leading to substantial increase of scanning time if not programmed in hardware or accelerated by other means.

\subsection{Practical limitations}
\label{practical_limitations}
FASTMER would provide the best results if the reference scan and the acquired scan are spatially aligned and exhibit a similar range of grey-level intensities. While these assumptions are mostly valid within the same imaging contrast (our slice similarity application), they may not be valid for different contrasts or scans acquired at different times. 

The solution to both issues can be obtained by grey-level normalization and realigning after acquisition. Since all data is acquired prior to reconstruction, a wavelet based recovery using all samples can be performed first. Although it may exhibit poor reconstruction of fine details (as presented in our experiments), it was found to be sufficient for grey-level normalization and alignment parameter extraction. Then, the extracted parameters are used for normalization and realignment of the data to improve reconstruction performance. An additional approach that is currently left for future research is to examine more complex similarity measures (as opposed to the grey-level subtraction proposed in this paper), e.g. non-linear similarity measures or measures that also take into account misalignement and intensity variation between the images.  However, one should take into account that the solution to the corresponding $\ell_1$ minimization problem with a non-linear constraint might be more complex than
FASTMER

It is worth noting here that if the alignment and normalization processes fail, then our iterative approach will detect the low similarity between the scans. As a result, the reference image will not be taken into account and reconstructed image will converge to a wavelet-based reconstruction. This statement is supported by experiments presented in Appendix \ref{AppC}.

\section{Conclusions}
\label{sec:conclusions}
In this paper we introduced a new framework fast MRI by exploiting a reference image (FASTMER). We developed an iterative reconstruction approach that supports cases in which similarity to the reference scan is not guaranteed. we observe the issue of embedding prior images in MRI reconstruction to accelerate acquisition and improve SNR has been examined in the past, in this paper the similarity to the reference image is learned during reconstruction. As a result, non-valid data is ignored in the reconstruction process, which enables the applicability of the method to a variety of MRI applications.

We demonstrate the performance of our framework in three clinical MRI applications: Reconstruction of noisy, single-contrast data, multi-contrast reconstruction and longitudinal reconstruction. Results exhibit significant improvement versus wavelet+TV based reconstruction and other MRI application-specific approaches.

Thanks to the existence of reference images in various clinical imaging scenarios, the proposed framework can play a major part in improving reconstruction in many MR applications. Future work will consist of applying the method to a wider range of medical imaging settings, such as low dose CT and fMRI, as well as exploring the combination of CS for parallel imaging and the proposed approach.

\section*{Acknowledgements} The authors wish to thank the Gilbert Israeli Neurofibromatosis Center (GINFC) for identifying, anonymizing and providing the MRI datasets, and performing the scanning experiments at various NEX values for the application that utilizes similarity between adjacent slices. This work was supported by the Ministry of Science, by the ISF I-CORE joint research center of the Technion and the Weizmann
Institute, Israel and by the European Union's Horizon 2020 research and innovation programme under grant agreement No. 646804-ERC-COG-BNYQ.

\noindent The authors have no relevant conflicts of interest to disclose.
\appendix
\section{Adaptation of Algorithm 1 and 2 for reference-based SNR improvement}
The SNR improvement in Section \ref{extention_snr} requires the solution of (\ref{eq12}) in an iterative manner. For this purpose, we define the weights update as follows:
\begin{align}
\begin{split}
 w_1^i&=\frac{1}{1+[|\mat{\Psi}_3\vec{\hat{x}}|]_i} \\ 
 w_2^i&=\frac{1}{1+[|\mat{B}\vec{\hat{x}}|]_i}.
\label{eq06}
\end{split}
\end{align}

Below we describe Algorithms 3 and 4, which are adaptations of Algorithms 1 and 2 to this setting, \noindent where $\Gamma_{\lambda \mu }(\vec{z})$ is defined in (\ref{eqA7}), $\mat{F}_3^*=\textrm{diag}([\mat{F}^*,\mat{F}^*,\mat{F}^*])$ and $\mat{\Psi}_3^*=\textrm{diag}([\mat{\Psi}^*,\mat{\Psi}^*,\mat{\Psi}^*]$.

\renewcommand{\algorithmicrequire}{\textbf{Input:}}
\renewcommand{\algorithmicensure}{\textbf{Output:}}
\begin{algorithm}[H]
\caption{FASTMER for SNR improvement}
\label{algo3}
 \begin{algorithmic} 
 \renewcommand{\algorithmicrequire}{\textbf{Input:}}
\renewcommand{\algorithmicensure}{\textbf{Output:}}
\REQUIRE \hspace{3mm} \\
Fully sampled k-space of $\vec{x}$: $\vec{z}$; Tuning constants: $\lambda_1,\lambda_2$\\
Number of k-space samples added at each iteration: $N_k$\\
Expected fidelity of measurements: $\mat{A}$\\ 

\ENSURE Estimated image: $\mat{\hat{x}}$
 \renewcommand{\algorithmicrequire}{\textbf{Initialize:}}
 \REQUIRE \hspace{3mm} \\
 $\mat{W}_1=\mat{I}$, $\mat{W}_2=\mat{I}$; 
\renewcommand{\algorithmicrequire}{\textbf{Reconstruction:}}
 \REQUIRE \hspace{1mm} \\
\WHILE{$\vec{z} \ne \emptyset$}
\STATE Move $N_k$ new samples to $\vec{y}$ from $\vec{z}$ according to distance from center of k-space.
\STATE {\bf Weighted reconstruction:} Estimate $\vec{\hat{x}}$ by solving (\ref{eq12})
\STATE {\bf Update weights: } Update $\mat{W}_1$ and $\mat{W}_2$ according to (\ref{eq06})\\
\ENDWHILE
\end{algorithmic}
\end{algorithm}

\begin{algorithm}[H]\caption{SFISTA algorithm for solving (\ref{eq12})}
\label{algo4}
\begin{algorithmic}
\REQUIRE \hspace{3mm} \\
k-space measurements: $\vec{y}$ \\
Sparsifying transform operator: $\mat{\Psi}_3$\\
Inverse sparsifying transform operator: $\mat{\Psi}_3^*$\\
Fourier operator: $\mat{F}_3$\\
Inverse Fourier operator: $\mat{F}_3^*$\\
Expected fidelity of measurements: $\mat{A}$\\ 
Tuning constants: $\lambda_1,\lambda_2,\mu$\\
An upper bound: $L\geq \|\mat{AF}_3\|_2^2+\frac{\|\mat{W_1\Psi}_3\|_2^2+\|\mat{W_2}\mat{B}\|_2^2}{\mu}$
\ENSURE Estimated image: $\mat{\hat{x}}$
\renewcommand{\algorithmicrequire}{\textbf{Initialize:}}
 \REQUIRE \hspace{3mm} \\
\STATE $\vec{x}_1=\vec{z}_2=\mat{F}_3^*\vec{y}$, $t_2=1$
\renewcommand{\algorithmicrequire}{\textbf{Iterations:}}
 \REQUIRE \hspace{3mm} \\
\STATE {\bf Step k:} $(k\geq 2)$ Compute
\STATE $\nabla f(\vec{z}_k)= \mat{A}^*(\mat{F}_3^*(\mat{A}(\mat{F_3}\vec{z}_k-\vec{y})))$
\STATE $\nabla g_{1\mu}(\mat{W}_1\mat{\Psi}_3\vec{x}_{k-1})=\frac{1}{\mu}\mat{W}_1\mat{\Psi^*}_3(\mat{W}_1\mat{\Psi}_3\vec{x}_{k-1}-\Gamma_{\lambda_1\mu}\left(\mat{W}_1\mat{\Psi}_3\vec{x}_{k-1})\right)$
\STATE $\nabla g_{2\mu}(\mat{W}_2\mat{B}\vec{x}_{k-1})=\frac{1}{\mu}\mat{W}_2\mat{B}(\mat{W}_2\mat{B}\vec{x}_{k-1}-\Gamma_{\lambda_2\mu}\left(\mat{W}_2\mat{B}\vec{x}_{k-1})\right)$
\STATE $\vec{x}_k=\vec{z}_k-\frac{1}{L}(\nabla f(\vec{z}_k)+\nabla g_{1\mu}(\mat{W}_1\mat{\Psi}_3\vec{x}_{k-1})+\nabla g_{2\mu}(\mat{W}_2\mat{B}\vec{x}_{k-1}))$
\STATE $t_{k+1}=\frac{1+\sqrt{1+4t_k^2}}{2}$
\STATE $\vec{z}_{k+1}=\vec{x}_k+\frac{t_k-1}{t_{k+1}}(\vec{x}_k-\vec{x}_{k-1})$
\end{algorithmic}
\end{algorithm}

\section{Similarity maps and resulting weighting matrices }
\label{AppB}
In this appendix we present the similarity maps between each gold-standard to the reference image and the resulting weighting matrices, $\mat{W}_1$ and $\mat{W}_2$ for the cases examined in this paper. Note that although in the paper $\vec{x}$ is defined as a vector and the weighting matrices are defined as diagonal matrices, $\mat{W}_1$ and $\mat{W}_2$ are represented in this section as non-diagonal matrices, used to weight $\vec{x}$  represented as a 2D matrix, for convenience. In addition, the difference images refer to the absolute normalized difference.

\begin{figure*}
\centering
T2-FLAIR experiment\\
\vspace{1mm}
\includegraphics[width=135pt]{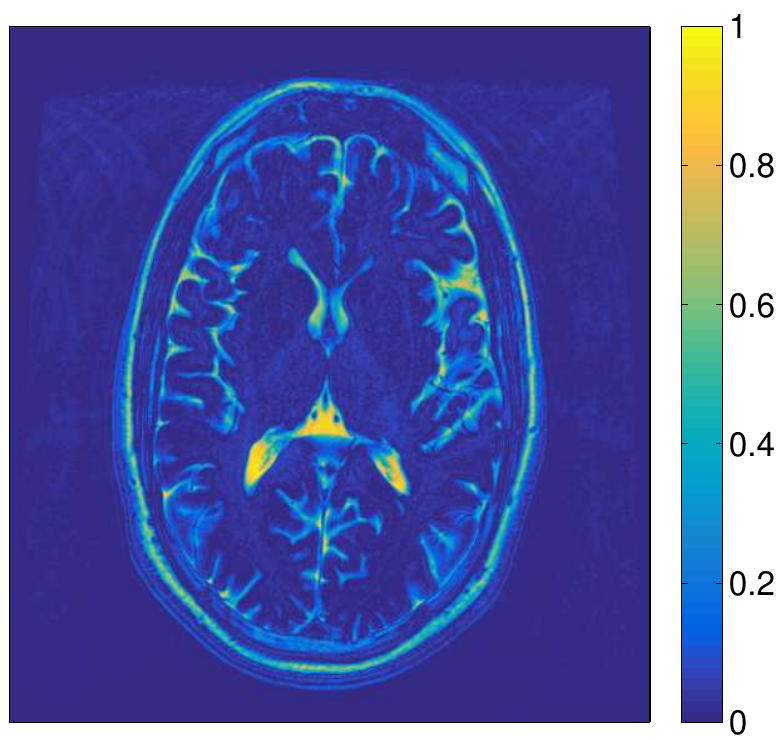}\hspace{8mm}
\includegraphics[width=135pt]{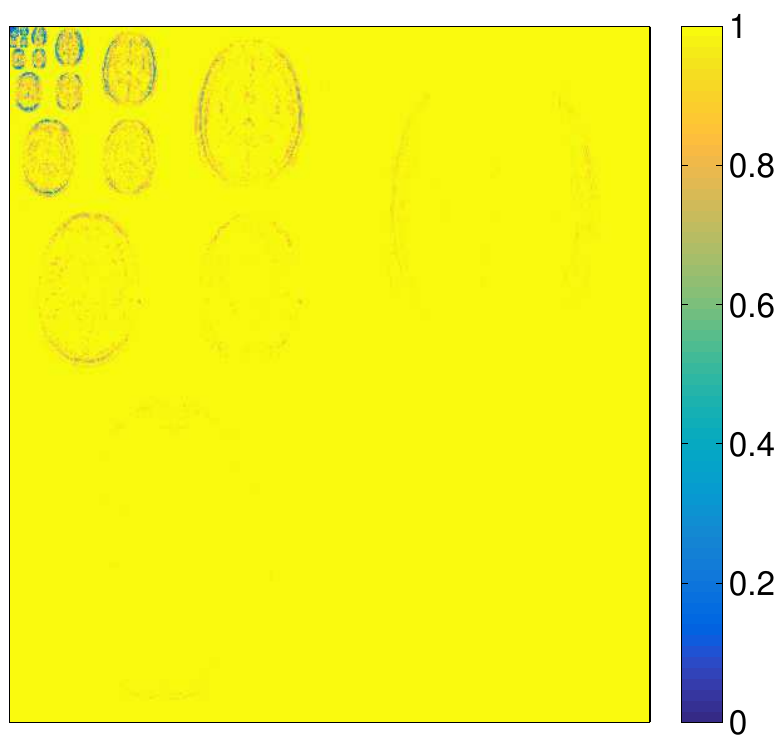}\hspace{8mm}
\includegraphics[width=135pt]{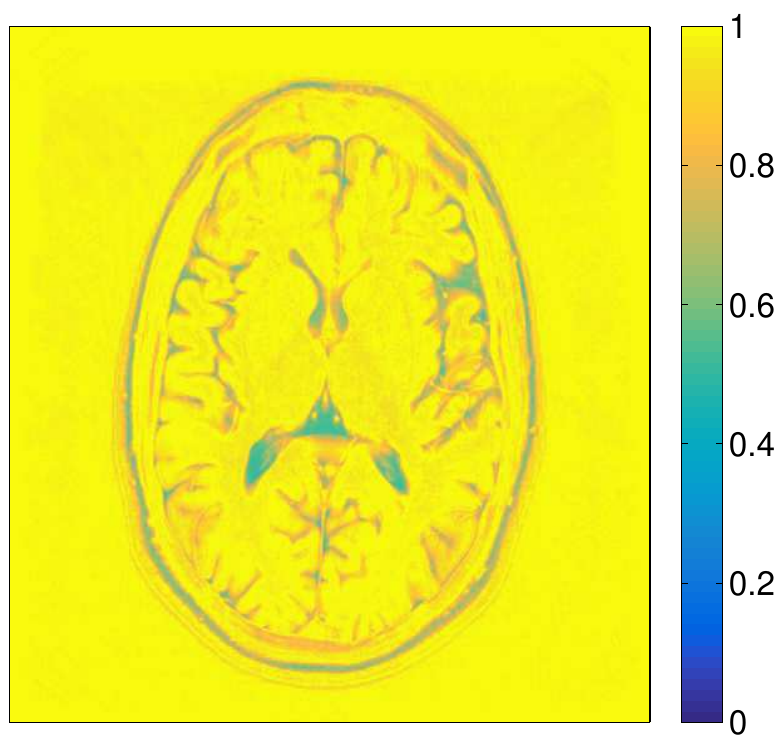}\\
\vspace{-5mm}{\footnotesize \begin{flushleft}\hspace{10mm}  Normalized difference image \hspace{30mm} Resulting $\mat{W}_1$\hspace{35mm}  Resulting $\mat{W}_2$ \end{flushleft}}
\caption{T2-FLAIR experiment: Normalized difference image between the gold-standard FLAIR image and the reference T2-weighted image, and the resulting weighting matrices, $\mat{W}_1$ and $\mat{W}_2$ of FASTMER.}
\label{appBfig1}
\end{figure*}

\begin{figure*}
\centering
Follow-up experiment\\
\vspace{1mm}
\includegraphics[width=135pt]{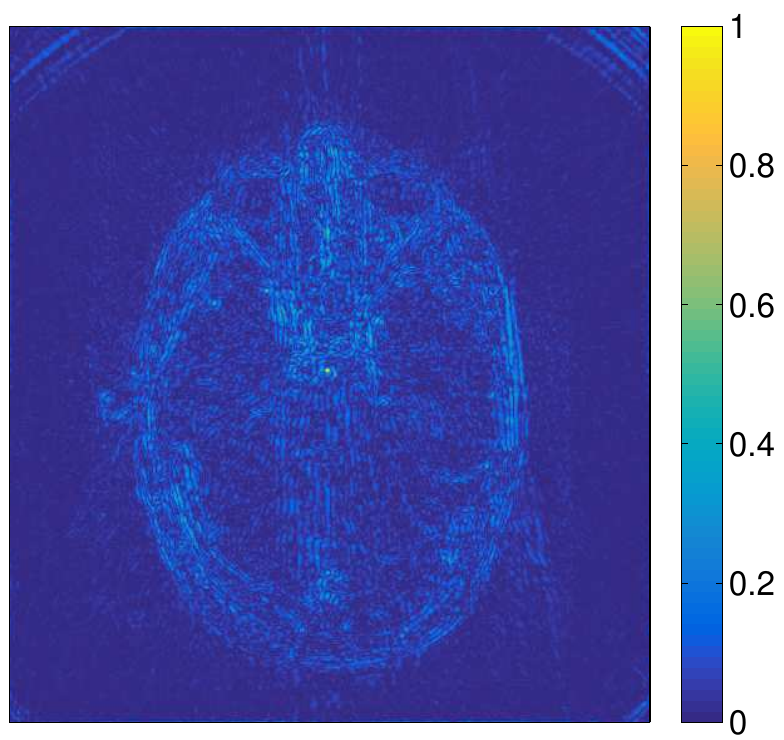}\hspace{8mm}
\includegraphics[width=135pt]{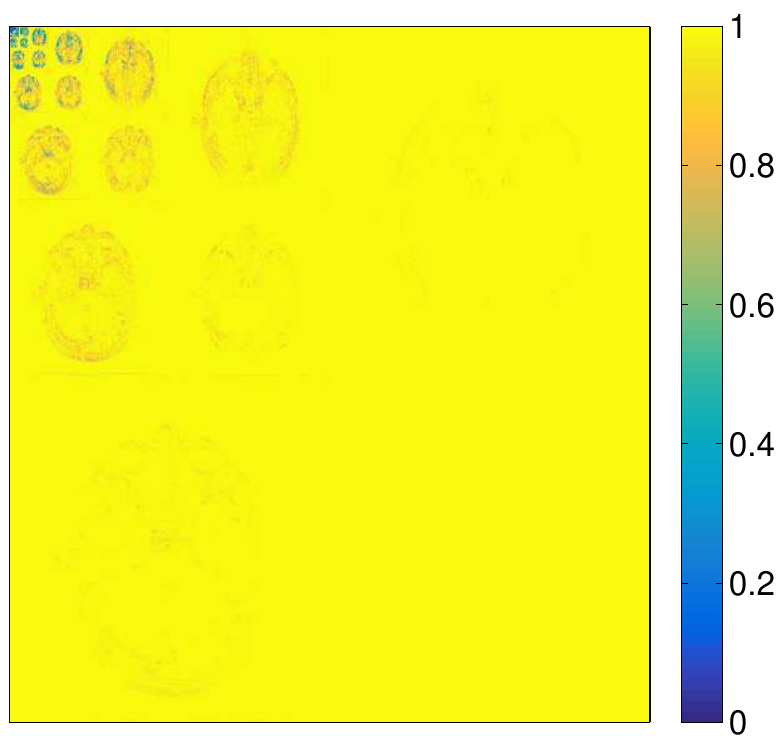}\hspace{8mm}
\includegraphics[width=135pt]{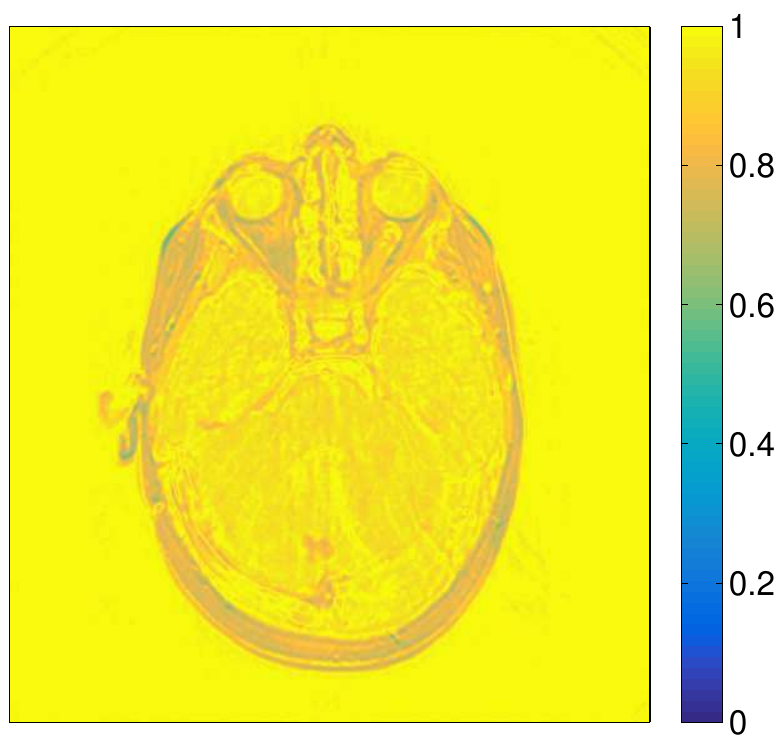}\\
\vspace{-5mm}{\footnotesize \begin{flushleft}\hspace{10mm}  Normalized difference image \hspace{30mm} Resulting $\mat{W}_1$\hspace{35mm}  Resulting $\mat{W}_2$ \end{flushleft}}
\caption{Follow up experiment: Normalized difference image between the gold-standard follow-up image and the reference baseline image, and the resulting weighting matrices, $\mat{W}_1$ and $\mat{W}_2$ of FASTMER.}
\label{appBfig2}
\end{figure*}

\begin{figure*}
\centering
SNR improvement experiment\\
\vspace{1mm}
\includegraphics[width=135pt]{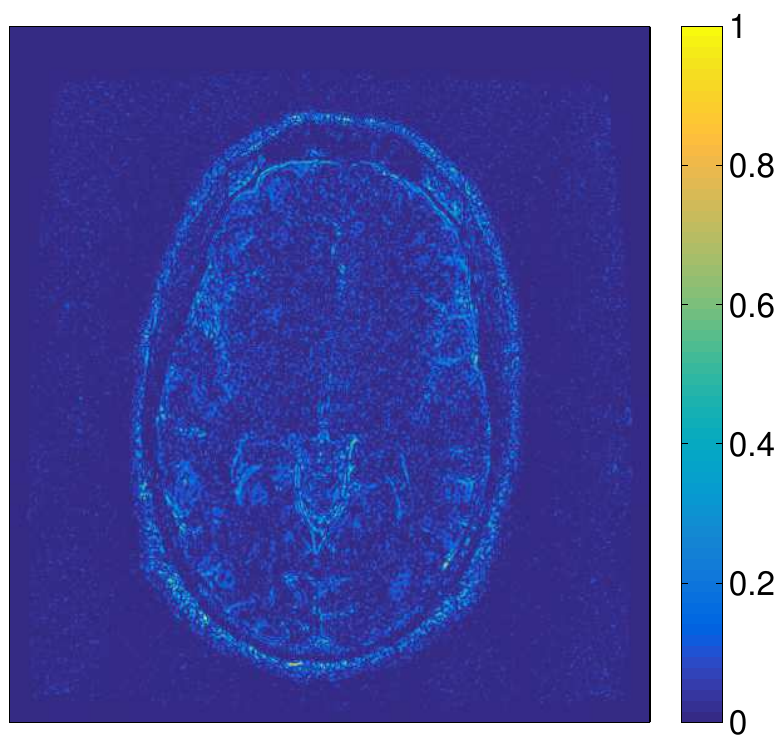}\hspace{8mm}
\includegraphics[width=135pt]{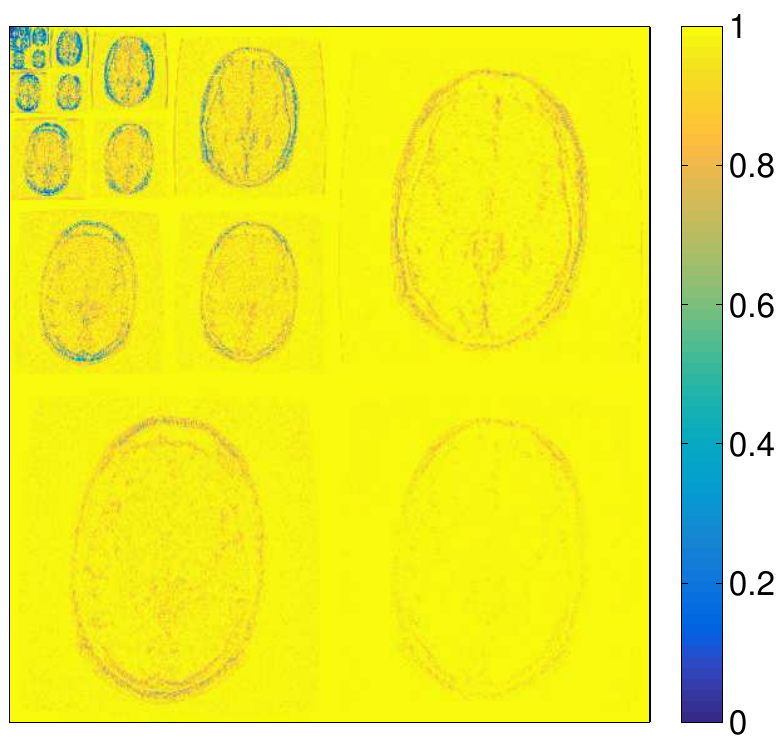}\hspace{8mm}
\includegraphics[width=135pt]{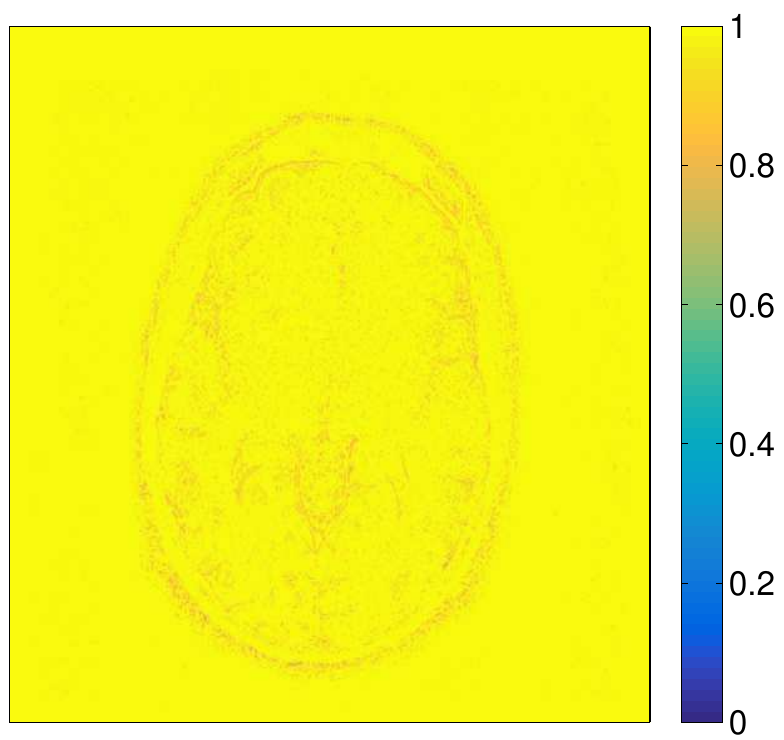}\\
\vspace{-5mm}{\footnotesize \begin{flushleft}\hspace{10mm}  Normalized difference image \hspace{30mm} Resulting $\mat{W}_1$\hspace{35mm}  Resulting $\mat{W}_2$ \end{flushleft}}
\caption{SNR improvement experiment: Normalized difference image between the two adjacent, low SNR input slices, and the resulting weighting matrices, $\mat{W}_1$ and $\mat{W}_2$ of FASTMER.}
\label{appBfig3}
\end{figure*}

Figures \ref{appBfig1} and \ref{appBfig2} show the maps for the T2-FLAIR and follow-up experiments and Figure \ref{appBfig3} shows the maps for the SNR improvement experiment. Here the difference map is computed between the input noisy adjacent slices.

\section{Method performance for varying degrees of similarity with the reference image}
\label{AppC}
To examine the performance of FASTMER for varying degrees of similarity between the reference image and the acquired image we repeated the experiment described in Section \ref{t2_flair_experiment} for misregistered reference image. Three scenarios were examined: 5\textdegree intra-plane rotation, 45\textdegree intra-plane rotation, and 5mm inter-plane translation together with 5\textdegree rotation. The results are shown in Table \ref{AppCfig1} , while the gold-standard, wavelet+TV and FASTMER results without mis-registration are given Fig. \ref{fig2}. 

It can be seen that the result of FASTMER converges to the result of TV+wavelet when a severe misregistration of 45\textdegree exists. This is indeed what was expected: FASTMER ignores the reference due to major changes between the reference and the acquired image. Rotation of 5\textdegree provides an improved reconstruction of PSNR=32dB, while a minor rotation and a 5mm translation provides results which are slightly better than the wavelet+TV reconstruction. The results of this experiment support the assumption described in the paper, that high degree of similarity with the reference image provides better results and vice versa. 
\begin{table*}[!ht]
     \begin{center}
     \caption{FASTMER results at various degrees of similarity with the reference image}
     \begin{tabular}{ |>{\centering\arraybackslash}m{3cm} | c | c | >{\centering\arraybackslash}m{2cm} | }
     \hline
Misregistration description
&Image used as a reference & FASTMER results &PSNR (dB) \\
 \hline
  \hline
\vspace{-40mm}  Rotation of 5\textdegree & \includegraphics[width=100pt,trim=1.9cm
 1cm 2cm 1.1cm, clip=true]{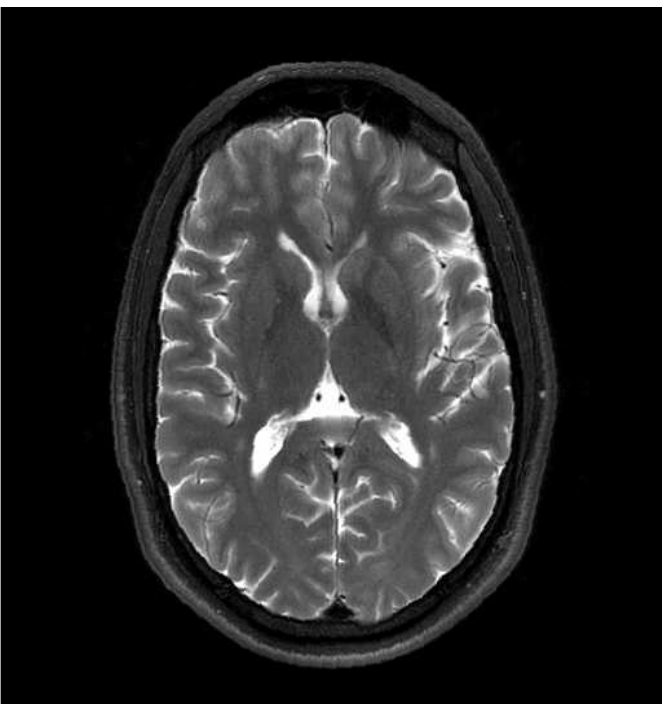} & 
\includegraphics[width=100pt,trim=1.9cm
 1cm 2cm 1.1cm, clip=true]{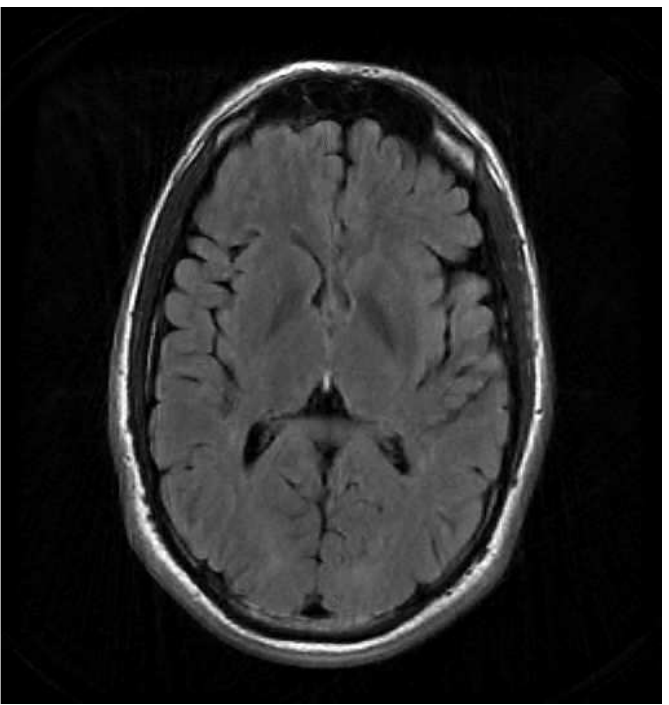} &
 \vspace{-40mm} 32 \\
 \hline
\vspace{-40mm}  Rotation of 45\textdegree & \includegraphics[width=120pt,trim=1.4cm
 1.1cm 1cm 1.35cm, clip=true]{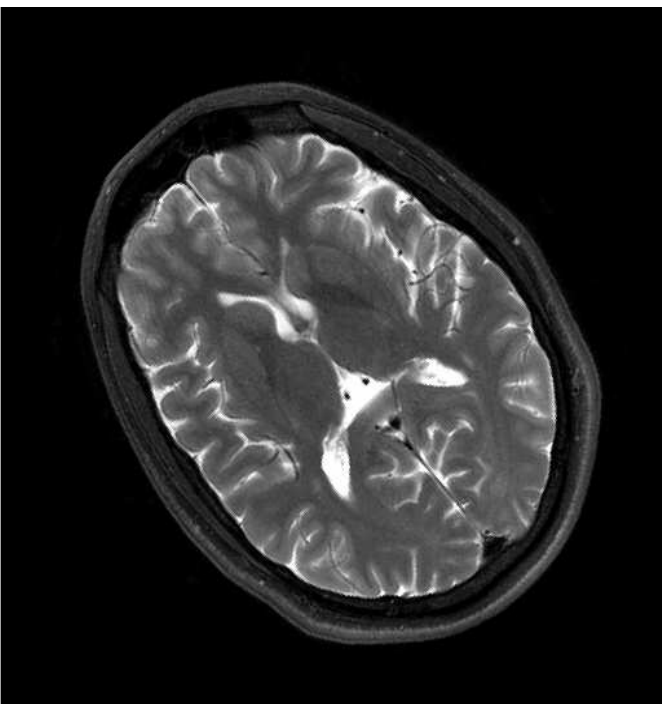}  & 
\includegraphics[width=100pt,trim=1.9cm
 1cm 2cm 1.1cm, clip=true]{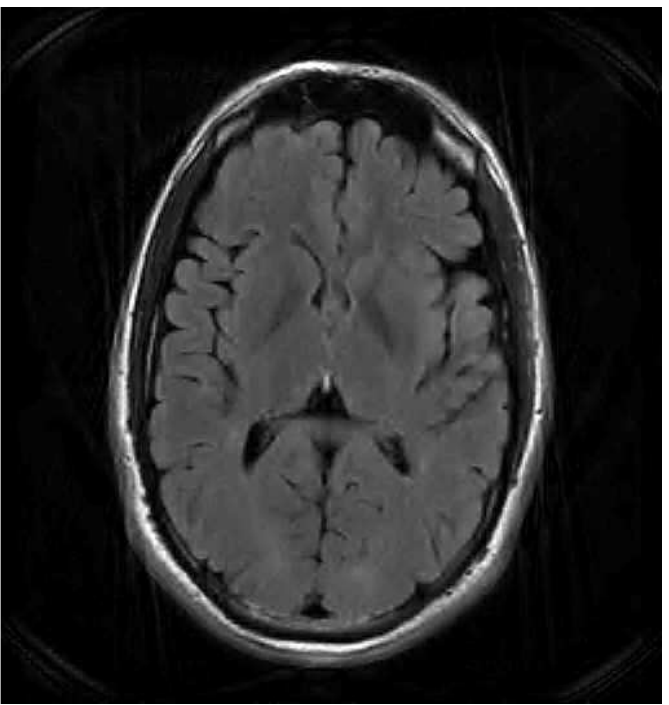} &
\vspace{-40mm} 28 \\
  \hline
\vspace{-40mm}  Rotation of 5\textdegree  + inter-plane translation of 5mm & \includegraphics[width=100pt,trim=2.2cm
 1cm 1.8cm 1.4cm, clip=true]{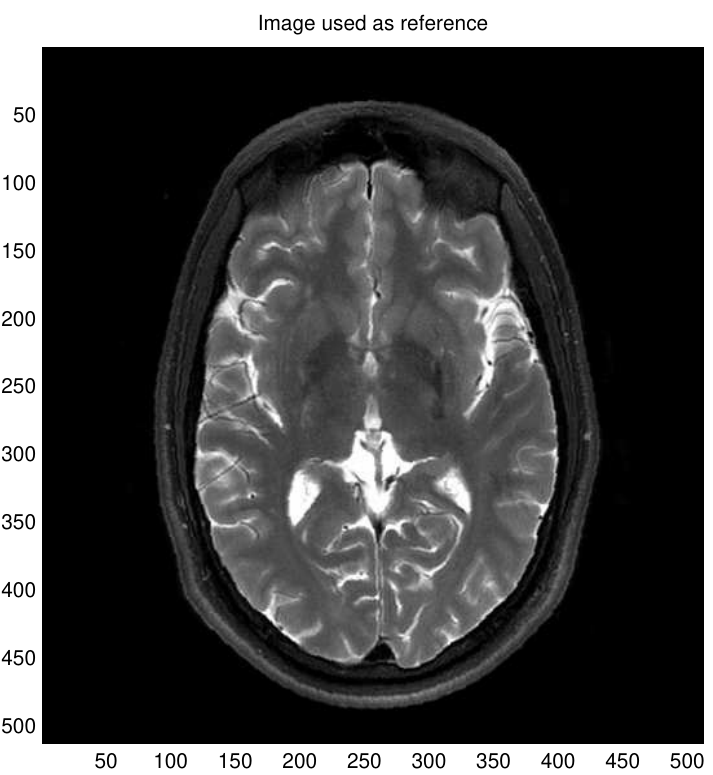} &\includegraphics[width=100pt,trim=1.9cm
 1cm 2cm 1.1cm, clip=true]{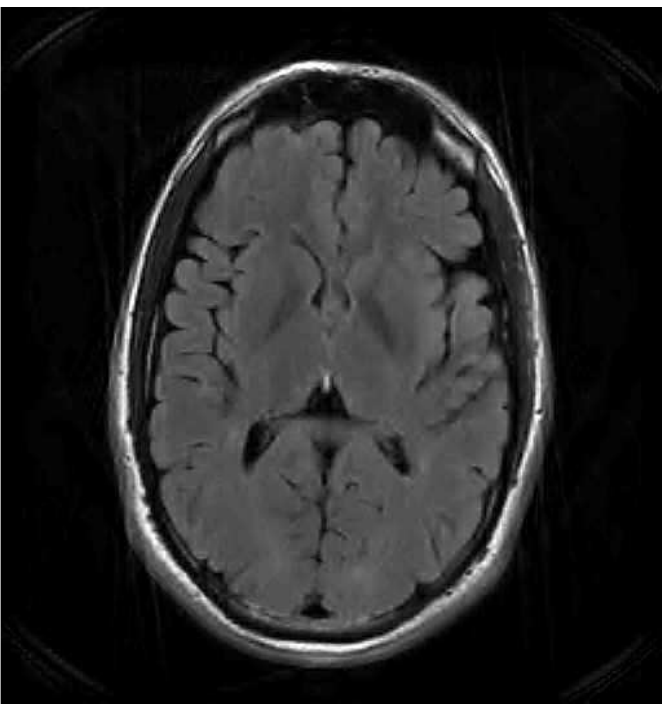} &
 \vspace{-40mm} 29 \\
  \hline
\label{AppCfig1}
\end{tabular}
     \end{center}
\end{table*}



\FloatBarrier
\bibliography{reference_based_mri_for_med_phys}

\end{document}